\documentclass[PRL, reprint,10pt,aps,amsmath,amssymb,longbibliography, superscriptaddress]{revtex4-1}

\usepackage{graphicx}
\usepackage{dcolumn}
\usepackage{bm}
\usepackage{hyperref}
\usepackage{lipsum}
\usepackage{comment}
\usepackage{nameref}

\begin{document}

\title{Machine learning and behavioral economics for personalized choice architecture}

\author{Emir Hrnjic}
\affiliation{CAMRI, NUS Business School 15 Kent Ridge Drive, Singapore, 119245}
\author{Nikodem Tomczak}
\email{nikodem.tomczak@nus.edu.sg}
\affiliation{Department of Chemistry, National University of Singapore, 3 Science Drive 3, Singapore, 117543}

\begin{abstract}
Behavioral economics changed the way we think about market participants and revolutionized policy-making by introducing the concept of choice architecture. However, even though effective on the level of a population, interventions from behavioral economics, nudges, are often characterized by weak generalisation as they struggle on the level of individuals. Recent developments in data science, artificial intelligence (AI) and machine learning (ML) have shown ability to alleviate some of the problems of weak generalisation by providing tools and methods that result in models with stronger predictive power. This paper aims to describe how ML and AI can work with behavioral economics to support and augment decision-making and inform policy decisions by designing personalized interventions, assuming that enough personalized traits and psychological variables can be sampled. 
\end{abstract}
\maketitle

\section{Introduction}
Incentives of economic agents lie in the center of economic reasoning, decision-making and policy-making \cite{neustadt1986thinking}. In fact, policy makers use the latest advancements in the understanding of incentives to design economic interventions in order to create a positive impact on the society. They often use one-size-fits-all solutions that provide very weak generalizations \cite{suh2019, coclanis2019, popper1986the}. Yet, economic interventions should be more effective by employing personalized incentives, i.e., on the level of individual members of the population \cite{halpern2016inside}.

The predictive power of policies based on economic theories that assume fully rational self-interested individuals vary broadly because humans often fail to behave like the fully rational \textit{homo economicus} \cite{levitt2008homo,thaler2015misbehaving}. In fact, human behavior and decisions deviate \cite{simon1955behavioral} from those of rational agents due to a multitude of psychological, cognitive, emotional, cultural, and social factors \cite{Tversky1974,kahneman2011thinking, wilkinson2012,just2014introduction}.

By exploiting these factors and selecting interventions to align incentives with certain objectives, policy makers can provide prescriptions to shape people's behavior \cite{thaler2008nudge} based on research in psychology and behavioral economics \cite{thaler2015misbehaving, thaler1980toward}. Choice architecture, the background against which people make choices, can include subtle nudges (hints, reminders or warnings) to drive human behavior toward "desired" actions while preserving the availability of choice through people's agency and control \cite{sunstein2017human}. Indeed, choice architecture has been successfully applied by policymakers \cite{gino2017, Chetty2015, congdon2011policy, madrian2014} targeting groups of people with certain commonalities \cite{sunstein2016, guszcza2015} resulting in numerous positive health care \cite{davenport2018}, economic, and societal outcomes \cite{halpern2016inside, benartzi2017}.

Nudging doesn't work very well on the level of individuals because of the poor generalisation of the economic incentives and the designed nudges. While the new policies result in a positive outcome on average (i.e., at the level of population), it is often not possible to tell whether and how effective they are for specific individuals.

Nudging a human collective on the level of individual members needs a different approach. Instead of asking how a particular nudge influences the behavior of a group of people we want the answer to a question how a particular individual would respond to a particular nudge given that individual's circumstances. 

We can't simply "brute-force" an outcome by trying many different policies on a population - this may still, on average, provide the same outcome by modifying the choices of a different sub-population. Neither can we provide a separate policy for each individual in society because of limited workforce and massive costs of such an undertaking.

What we need instead is a generalized model, which, when applied to each individual separately based on each individual's unique circumstances and context, will provide a prediction for most suitable personalized response that would yield the desired outcome.

Prediction methods that generalize and scale over large heterogeneous populations are being developed within the fields of data science, artificial intelligence, and machine learning.

The field of artificial intelligence (AI) concerns with the design and engineering of artificial, autonomous, rational agents \cite{russell2016,parkes2015} that can learn, make predictions and decisions, and improve them without human intervention. Machine learning (ML), a subset of AI, centers on iteratively learning algorithms in order to obtain generalized models which describe, often hidden, relationships within and between datasets \cite{goodfellow2016deep, hastie2009the}. The obtained models are then tested against known outcomes, and, if satisfactory, are used to make predictions on out-of-sample data. The models are improved with each new input-output query through the application of appropriate feedback to the model-generating algorithm.

Artificial intelligence and machine learning originated in the middle of last century \cite{nilsson2010the,simon1996the} and have seen a resurgence in popularity in recent years primarily driven by the availability of data \cite{einav2014}, reduced computing costs, and progress in mathematical algorithms \cite{provost2013data,jordan2015,lecun2015}. Importantly, making predictions became affordable; simple ML models can be trained and deployed on a personal laptop. Moreover, complex ML algorithms requiring dedicated hardware are well within reach of most corporations and organizations.

The applications of machine learning range from natural language processing \cite{grimmer2013}, medical diagnostics \cite{wang2016}, predictive business analytics \cite{provost2013data, bajari2015machine, Varian2014}, persuasion profiling \cite{pariser2011} to predicting crime and recidivism \cite{Dressel2018, chandler2011predicting}, and distributions of wealth \cite{blumenstock2015predicting, blumenstock2016, Jean2016}, among many others \cite{Brynjolfsson2017}. Machine learning and AI are used in the development of autonomous self-driving cars which continuously scan the surroundings using cameras and sensors, identify objects and patterns in the data they collect, predict behaviors of traffic participants and make decisions based on these predictions; all without human intervention \cite{bonnefon2016social,gerrish2018how}. 

In banking and finance, machine learning is used for credit-scoring, predicting customer attrition, facilitating loan decisions \cite{huang2007credit}, trading \cite{wellman2017ethical, deng2016deep}, advisory, and customer service, among others \cite{metz2016}. Applying ML to financial predictions allows for the detection of interactions in the data that are invisible to existing financial economic theories \cite{heaton2017deep}. 

Machine learning has been recently applied in the fields of economics \cite{parkes2015, mullainathan2017machine, Varian2014}, psychology \cite{joel2017romantic} as well as physical and chemical sciences aiding and accelerating discovery of new materials \cite{carrasquilla2017machine,butler2018,lookman2019active}, finding hidden patterns in large datasets \cite{biamonte2017quantum, Ramprasad2017, oganov2019structure}, and planning chemical synthesis \cite{segler2018planning}. In future, AI and ML will, at least to some degree, replace parts of the scientific discovery and innovation processes through automation and unsupervised learning \cite{extance2018}.

In short, ML can find previously obscured nonlinear patterns in data and use that to make more accurate predictions in various fields. In the next section, we focus on its ability to predict human behavior.

\section{From prediction machines to choice architecture machines}

We can use the machine learning toolbox to significantly improve human decision making \cite{kahneman2016,agrawal2018prediction} and improve our ability to predict what person will respond to what persuasive technique \cite{chandler2011predicting, rosenfeld2012combining, Risdon2017,subrahmanian2017predicting, ascarza2018retention, andini2018targeting, plonsky2019predicting}. However, even when using predictions obtained from ML methods users still have to form their own judgement \cite{loewenstein2015}. In this context, ML may augment users' decision-making process by providing possible solutions, which they may have missed otherwise. In contrast to computers, peoples' ability to process thousands of variables in a short time is limited and depends on their emotional state. ML reduces the time it takes to make a decision and the cognitive cost of judgement \cite{agrawal2018prediction}. ML can therefore prompt to reconsider previously made decisions and to carefully reevaluate the possible outcomes.

\section{General formulation}

We present a general formulation of the problem where a decision of the choice of a personalized intervention is augmented by a prediction using machine learning models. 

Assume $G$ to be the utility describing an \textit{a priori} specified goal \cite{mitchell2018prediction}. Let $G^{\Omega}(N_{u})$ be the total utility of our decision, made over a populations $\Omega$, to apply nudge $N _{u}$ selected from a set of nudges $N$, where $u = 0,...,l$ and $N_{0}$ means no nudging. By selecting only one nudge that would selectively influence a subset of $\Omega$, utility $G^{\Omega}$ may be very far from its maximum value. If we change $N _{u}$ we obtain a different $G^{\Omega}$. We may however not know how different $N _{u}$ affects individual members of the population.

Consider that the outcomes of our decisions can be evaluated at the level of individuals. To maximize outcome $G^{\Omega}$ for a problem $P$, we adopt a personalized intervention approach, where for every person $\omega _{i} \in \Omega$, $i = 1,...,n$ , we can assign $N _{u}$, such that we maximize the utility $G^{\omega _{i}}(\delta^{u} _{i})$, where $\delta^{u}_{i}$ is a decision to use a nudge $N_{u}$ for person $i$. 

To obtain a prediction model, for each $\omega _{i}$ we assemble a feature vector $\Pi _{i}$, which contains a set of informative variables, traits, $\pi _{k} \in \Pi$, where $k = 1,...,m$ denotes the individual variable in the set, to which we add a variable $\pi _{m+1} = N _{u}$ corresponding to a particular nudge, and an outcome variable $y _{i}: \left \{0,1\right\}$, i.e., we assign "1" or "0" depending on whether the nudge worked for a particular person or did not, respectively. The prediction model, or hypothesis, is obtained by supervised learning by training on an experimentally obtained dataset consisting of many input-output pairs ($\Pi _{i}\cup \left \{N _{u}\right\}, y _{i}$). 

Predicting best $\delta _{i}^{u}$ for people which were not in the training set is then made by estimating the probabilities $P\left ( y_{i}= N_{u}\mid\Pi_{i}\right )$. Our decision can then be defined as
\begin{equation}
\delta _{i}^{u} = \underset{u}{\arg\max} P\left ( y_{i}= N_{u}\mid\Pi_{i}\right )
\end{equation}

To maximize the total utility we need to sum over all the individual utilities  $G^{\omega _{i}}(\delta^{u}_{i})$:
\begin{equation}
\max G^\Omega = \sum_{i=1}^{n}G^{\omega _{i}}(\delta^{u}_{i})
\end{equation}

Applying a particular nudge incurs costs. These costs are summative - if more nudges are applied, the cost is the sum of the cost of individual nudges. The payout of nudges is commutative, i.e., it doesn't matter in which sequence they are applied, however it is also non-associative (applying two nudges in sequence may not be the same as applying two at the same time). Applying more nudges then necessary for a single individual at the same time will therefore have diminishing returns.

The question is then whether applying personalized nudges is always better than doing nothing, i.e., whether a wrongly prescribed nudge can be worse than no nudging and whether there are externalities \cite{mitchell2018prediction, athey2017beyond}. 

Not nudging is included in the set $N$. The model may predict that we will be better off without prescribing any intervention, i.e., $P\left (N_{0}\mid\Pi_{i}\right )$ will be larger than for any other nudge. We can also run into a situation where the probabilities for every outcome will be very low. In this case we may wish to estimate a cut-off probability below which we would default to not nudging \cite{loewenstein2015}. This cut-off may be related to the cost of nudging and may be important at the early stages of model deployment when the model may give poor predictions. Of note is that personalized nudging may at the end significantly cut cost by avoiding wrongly prescribed, or effectively ineffective, nudges.

Finally, we need to evaluate whether the utility of personalized nudging is higher or equal than the utility of the one-nudge-for-all approach for every possible nudge.

\begin{equation}
\forall N_{u} \in N,  G^\Omega(N_{u}) \leqslant G^\Omega(\delta^{u})
\end{equation}

Such situation may occur as the ML models are not 100\% accurate and can assign wrong nudges \textit{en masse}, especially at the early stages of model employment increasing the cost, compromising fairness and drastically decreasing the total utility of personalized predictions.

\section{Experimental approach}

\subsection{Populations and feature selection}

The populations over which choice architects would want to make personalized decisions are usually not randomly selected, but depend on the problem that needs to be solved. For example, post treatment compliance in diabetics will necessarily reduce the population to people that are already sick and with traits uniquely characterizing them, such as, perhaps, a sedentary life style. The designed set of nudges $N$ will therefore be biased and depend on the sample. There is however enormous amount of research data from behavioural studies available that could provide a meaningful hint which variables may be relevant \cite{benartzi2017}. To deeply understand which nudges should be included into $N$, choice architects need to better understand the problem by running interviews with practitioners and reviewing literature describing results from single nudges. Such approach will also help in feature selection. Identifying and removing co-linear variables through feature selection is essential to minimize overfitting.

Random forests algorithms \cite{breiman2001random} bootstrap subsamples of predictors and handle many predictors at once while minimizing overfitting. At the same time, they are sensitive to nonlinear relationships and to complex interactions between predictors. However, unlike single regression trees where the relationship between the data is much clearer, random forest algorithms are harder to interpret. Transparent and easily communicable algorithms may take preference in personalized prediction of human behavior due to ethical considerations even at the expense of the overall intervention cost due to less accurate predictive models.

The complexity of human traits and human behavior may generate numerous outliers in datasets. However, these outliers should not be removed in personalized interventions, since choice architects prefer to provide a solution for everyone. Choice architects need to account for them in order to design models that will not only work for the centre of the distributions.

Using preregistration, akin to the one used in clinical trials and psychology, will make the experiments more transparent, less prone to error, and easier to reproduce \cite{kupferschmidt2018recipe, Hutson2018}. Preregistration may include the methods and predictors that will be used to train the ML models.

Reinforcement Learning is a machine learning method that helps an agent learn from experience by maximizing a cumulative reward. Although reinforcement learning doesn't need much historical data in advance it can take long time to learn. It could be used to complement individualized response models with learning from population outcomes. However, it needs to be deployed for the learning to start. In this case, one needs to carefully evaluate the cost of wrongly prescribed nudges. If the cost is high, bootstrapping the models with available literature data may be advantageous.

\subsection{Model performance testing}

Evaluation metrics such as accuracy, confusion matrix, logarithmic loss, or F-score which includes precision and recall, and their modifications, are often used to evaluate classification models on out-of-sample datasets. In context of decisions that affect people directly, the choice of performance metrics to optimize depends however on the problem we try to solve and on our measure of fairness \cite{kusner2017counterfactual,zemel2013, Kleinberg2018b}. The later needs to ideally be defined a priori. Different stakeholders may see the usefulness of the metrics differently, but, in context of nudges, support for human values should take priority to mathematical correctness. The metrics also may assume that the test set is representative of the underlying real world task and that the model's logic is free of errors, i.e., that the model is neither accurate for the wrong reasons \cite{mudrakarta2018did} nor oversensitive to certain details.  

In practice we may wish to evaluate the model performance at the level of the population, groups with the same interventions allocated, and on the level of individuals to see whether predictive value parity takes undue precedence over other metrics \cite{zemel2013, calmon2017}. Accuracy alone will therefore seldom be the right evaluation metrics. 

A crucial aspect of using ML models for predictions is to continuously improve their performance in a ''learning-by-using'' fashion. In some cases, this can be achieved via interviews. It would be advantageous to know why certain nudge did not work to eliminate the likelihood of anomalies entering the model learning loop.

While the model is deployed, one should also continuously test its efficacy due to numerous reasons. The first reason is model decay causing the model to slowly drift towards less generalized solutions. This can be due to concept drift, i.e., changes how we interpret the data, but also due to previously unseen variety of data and differences in the meaning of labels which may cause a shift of decision boundaries. The second reason is nudge creep, i.e., the idea that people who are nudged and aware of it will adapt their behavior neutralizing the effect of the nudge. The third reason is the efficacy of the nudge. For example, weight loss incentives may work only for obese people with a particular lifestyle or eating habits. If the nudge worked and the person actually lost weight, presumably due to a change in lifestyle or eating habits, the same nudge applied again for the same person may be much less effective. Finally, it would be important to collect and include personal historical data for improved model training.  

\section{Challenges}
\subsection{Data availability and augmentation}

The success of applying machine learning approaches hinges on the availability of reliable and bias-free data which contain informative variables. Depending on the ML task, the amount of required data can range from hundreds to millions of records. To ascertain outcomes with confidence, typically large scale and high resolution data is needed. The necessary dimensionality of the required data to make personalized prediction is however not entirely known. Increase in data dimensionality does not automatically result in better predictions and some conceptual insight into which variables would be most predictive is necessary \cite{buchanan2019}.

Some data are already available to different stakeholders - for example, socioeconomic data \cite{einav2014} to the public, food purchases \cite{aiello2019} to stores, medical prescription records to doctors, healthcare records to hospitals, etc. 

Randomized control trials, where subjects are randomly allocated different intervention and there is one control group that has not been nudged have been widely used to devise best nudging strategy. The data from these studies, if publicly available, would be ideal for the personalized intervention as described here \cite{halpern2016inside}. The trials come with their own set of challenges and biases that need to be evaluated \cite{potash2018}. Alternatively survey data can be used as well.

If available data is limited and its use results in models with low predictive power, one may also consider data augmentation. In data augmentation small variations to the existing dataset are introduced through data manipulation. For example, distorting, or adding noise to a picture have been shown to overall help obtaining better image recognition models. Although such approach works very well for some tasks such as image recognition, applying data augmentation to describe human behavior is not straightforward. It is not clear which features, and to what extent, could be manipulated and whether such manipulation is free from ethical concerns. Application of data augmentation to machine learning problems that analyze human behavior and provide a basis for decision making that is expected to impact people directly clearly warrants consideration and careful research studies.

\subsection{Ethics of AI, model interpretability and unintended consequences}\label{subsec:ethical}

Measures of fairness of ML algorithms and resulting models are intensively debated and understanding the trade-offs associated with each employed measure is crucial for making informed decisions \cite{angwin2016, kleinberg2018}. Application of machine learning algorithms to data that is finite and inherently biased may amplify the discriminatory biases \cite{mitchell2018prediction, silberg2019} mirroring the societal stereotypes. Removing these biases from the data may be arduous, and workarounds would need to be devised \cite{kleinberg2017, dwork2018} if we want to have a technology that will move us to a more equitable society for all \cite{oneil2016}. Nondiscrimination and respect for fundamental human rights must be preserved through the preservation of agency and control regardless of incentives and economic forces - we should be free from unwanted manipulation and able to exercise our free will at all times \cite{sunstein2016}. 

Jon Kleinberg et al. \cite{kleinberg2016,kleinberg2017} showed that in some cases data science and machine learning can, based on populations data, find behavioral-type variables that can be used to identify biases and help predicting outcomes \cite{kleinberg2017} on the level of individuals. In other cases, we may need to sacrifice utility to achieve higher fairness.

We need to interpret, understand, and explain the solutions prescribed by ML models \cite{Dressel2018,Voosen2017, ribeiro2016, molnar2019} to ascertain the consequences of misaligned incentives. Even if our "black box" tool makes better predictions compared to practitioners we still are responsible for the prediction's unintended consequences and minimizing harm should be priority \cite{Rahwan2019}. Frameworks like Local Interpretable Model-Agnostic Explanations (LIME) \cite{ribeiro2016} can provide insights into how the machine learning models make decision in a way that is readable for human non-experts. In context of individualized recommendations, the main challenge will be the availability of interpretable representations of individualized predictions, instead of the entire model behavior. This challenge needs to be addressed to keep improving the models \cite{Johansson2011} and make them trustworthy to the nudged subjects and the policymakers.

In context of personalized intervention, a wrongly prescribed nudge must in general not be worse than no nudging at all - the counterfactual. Nudges should also be fair - we should genuinely try to solve a problem regardless of people's circumstances \cite{mitchell2018prediction}. A tight collaboration between computer scientists and stakeholders with knowledge about the problem may be needed to move beyond simple predictions to causal inference \cite{athey2017beyond,subrahmanian2017predicting, varian2016causal} that would allow to build better models. Due to cultural and demographic variations, machines and machine decisions may shape human behavior in unintended ways based on where and how they are deployed creating new social problems as a result \cite{awad2018}.

A possible large negative impact of precision nudging may be unethical influencing and harnessing of collective behaviors by applying individualized impulses (nudges) to a specific subset of a target population \cite{frischmann2018,frischmann2018b}. This could lead to swarm behavior similar to that observed in animals \cite{ried2019modelling}, as well as herding behavior among humans under specific conditions \cite{sumpter2005principles,schelling1978micromotives}. To mitigate this, the possibility of adversarial attacks on the machine learning models would need to be researched and evaluated carefully \cite{mudrakarta2018did,Finlayson1287, Kurakin2016} to prevent interference by third parties with competing incentives.

\section{Precision nudging applications}

It is important for experimental economists to ascertain whether a behavior inside the laboratory is a good indicator of behavior outside the laboratory \cite{levitt2007, Falk2009, Charness2015}. Ultimately, one needs models that generalize well to the real world \cite{levitt2008homo}. In natural circumstances, a person's behavior and decisions will be based on a complex combination of personality traits and situational context. Some nudges may even have adverse effect as has been clearly demonstrated experimentally in the past \cite{halpern2016inside}. In fact, based on these findings it has been suggested that population segmentation based on a single variable, such as personal wealth, may already boost the efficiency of certain interventions. Conversely, applying personalized interventions based on predictions made by ML models could increase the total utility of our decisions. 

Below we list selected examples how precision nudging may be applied. This list is definitely not exhaustive.

\subsection{Mobile user behaviour}

Microtargetting and allocating interventions has been used in mobile applications for many years by following each user behavior over time, including group behaviour of users with certain commonalities \cite{downs2013}. Predicting what the user may do next, interventions are typically applied even before the user makes a decision. This could be pushed one step farther - by analyzing many users' behavior the nudges could be selected with much better precision \cite{ascarza2018retention}.

\subsection{Corporate policy}

A large firm conducted field experiments to measure the effect of prompts on influenza vaccination receipts offered to its employees at free on-site clinics. The discerned prompts resulted in a statistically different vaccination rates \cite{Milkman2011}. Machine learning could potentially be used to predict which prompt would work best for which employee and compare the "precisely nudged" employee vaccination rate to the control values obtained from the study. 

\subsection{Nutrition}

While the idea of an universal diet is being superseded by personalized nutrition designed to induce individualized response to food based on our genetic makeup, physiology, habits, gut microbiome etc. \cite{Zeevi2015}, of equal importance is to make sure that the prescribed food will be actually consumed. To this end, personalized nudges would need to be devised such that a subset of foods will be consumed by a specific person. General nudging has already been shown to work when targeting populations dietary or health habits, but its effectiveness on the level of individuals is rarely known.

\subsection{Healthcare}

The majority of hospital re-admissions after surgery are due to non-adherence to discharge protocols. If the patient does not adhere to the prescribed treatment outside the hospital, the overall medical intervention is seriously compromised. Machine learning models that would predict which precision nudges would most effectively serve discharged patients would significantly reduce this cost. Since readmission data are available, it is tempting to correlate different types of nudging employed by different healthcare providers with the readmission rates and see whether the patients' data can provide clues on the willingness of a particular patient to follow different clinicians’ recommendations, i.e., the nudges.

To reduce healthcare cost, each person could be nudged in a different way to adopt a set of health behaviors tailored to their circumstances, place of residence, health record, social conventions, and personal biases, among other variables. The challenge is to find out which data would provide the most meaning and which variables will be a strong predictive feature for the machine learning models to design the personalized nudge. 

\subsection{Conservationism}

Nature conservationists want to utilize a behavioral science toolkit to address unsolved complex wildlife conservation issues \cite{Park2019}. Conservation is a behavioral problem and a catalogue of nudges drawn from behavioral economics has been proposed, but not yet tested. Whether the target behavior has been achieved using a particular nudge will be very hard to asses, but it can provide a wealth of data for ML. As conservation is a complex issue, nudging self-interested individuals may be a good strategy.

\section{Conclusions}

Precision nudging is a new field of research that combines behavioral economics principles with machine learning and artificial intelligence methods. It will influence how we make decisions and how we design public policies; it is libertarian paternalism \cite{thaler2003, camerer2003} taken to extreme. Instead of looking for one-size-fits-all solutions, one should look for models that generalize over each and every member of the targeted population. Such approach would likely need to rephrase the questions typically asked in choice architecture studies and would need to rely on large and reliable datasets. Digitization of public records and the ubiquity of data collection would help. In some circumstances, surveys and experiments can be designed to obtain appropriate data. As behaviour predicts behaviour, access to time dependent variables may be very valuable. Personalized interventions may remove biases inherent to our decision making. 
Combining machine learning with behavioral economics would allow economists to run experiments and validate hypotheses regardless of the underlying choice mechanism. Such experiments should be background agnostic and tractable - what is important is whether the machine learning models generalize well enough to provide for a personalized intervention. If this can be achieved, precision nudging can become a General Purpose Technology \cite{bresnahan1992} that could be applied to many problems we face today. If proven to be widely effective, precision nudges will be adopted to augment human decisions in workflows in policymaking, finance, healthcare, education, and many other fields.

\section{Authors' Contributions}
NT conceived and planned the research. EH and NT wrote the manuscript. 


\begin{thebibliography}{114}%
\makeatletter
\providecommand \@ifxundefined [1]{%
 \@ifx{#1\undefined}
}%
\providecommand \@ifnum [1]{%
 \ifnum #1\expandafter \@firstoftwo
 \else \expandafter \@secondoftwo
 \fi
}%
\providecommand \@ifx [1]{%
 \ifx #1\expandafter \@firstoftwo
 \else \expandafter \@secondoftwo
 \fi
}%
\providecommand \natexlab [1]{#1}%
\providecommand \enquote  [1]{``#1''}%
\providecommand \bibnamefont  [1]{#1}%
\providecommand \bibfnamefont [1]{#1}%
\providecommand \citenamefont [1]{#1}%
\providecommand \href@noop [0]{\@secondoftwo}%
\providecommand \href [0]{\begingroup \@sanitize@url \@href}%
\providecommand \@href[1]{\@@startlink{#1}\@@href}%
\providecommand \@@href[1]{\endgroup#1\@@endlink}%
\providecommand \@sanitize@url [0]{\catcode `\\12\catcode `\$12\catcode
  `\&12\catcode `\#12\catcode `\^12\catcode `\_12\catcode `\%12\relax}%
\providecommand \@@startlink[1]{}%
\providecommand \@@endlink[0]{}%
\providecommand \url  [0]{\begingroup\@sanitize@url \@url }%
\providecommand \@url [1]{\endgroup\@href {#1}{\urlprefix }}%
\providecommand \urlprefix  [0]{URL }%
\providecommand \Eprint [0]{\href }%
\providecommand \doibase [0]{http://dx.doi.org/}%
\providecommand \selectlanguage [0]{\@gobble}%
\providecommand \bibinfo  [0]{\@secondoftwo}%
\providecommand \bibfield  [0]{\@secondoftwo}%
\providecommand \translation [1]{[#1]}%
\providecommand \BibitemOpen [0]{}%
\providecommand \bibitemStop [0]{}%
\providecommand \bibitemNoStop [0]{.\EOS\space}%
\providecommand \EOS [0]{\spacefactor3000\relax}%
\providecommand \BibitemShut  [1]{\csname bibitem#1\endcsname}%
\let\auto@bib@innerbib\@empty
\bibitem [{\citenamefont {Neustadt}\ and\ \citenamefont
  {May}(1986)}]{neustadt1986thinking}%
  \BibitemOpen
  \bibfield  {author} {\bibinfo {author} {\bibfnamefont {Richard}\ \bibnamefont
  {Neustadt}}\ and\ \bibinfo {author} {\bibfnamefont {Ernest}\ \bibnamefont
  {May}},\ }\href@noop {} {\emph {\bibinfo {title} {Thinking in time: the uses
  of history for decision-makers}}}\ (\bibinfo  {publisher} {Free Press Collier
  Macmillan},\ \bibinfo {address} {New York London},\ \bibinfo {year}
  {1986})\BibitemShut {NoStop}%
\bibitem [{\citenamefont {Suh}(2019)}]{suh2019}%
  \BibitemOpen
  \bibfield  {author} {\bibinfo {author} {\bibfnamefont {Bob}\ \bibnamefont
  {Suh}},\ }\bibfield  {title} {\enquote {\bibinfo {title} {{Sales Teams
  Aren’t Great at Forecasting. Here’s How to Fix That}},}\ }\href@noop {}
  {\bibfield  {journal} {\bibinfo  {journal} {Harvard Business Review}\ }
  (\bibinfo {year} {2019})},\ \bibinfo {note} {accessed 2019-05-19}\BibitemShut
  {NoStop}%
\bibitem [{\citenamefont {Coclanis}(2019)}]{coclanis2019}%
  \BibitemOpen
  \bibfield  {author} {\bibinfo {author} {\bibfnamefont {Peter~A}\ \bibnamefont
  {Coclanis}},\ }\bibfield  {title} {\enquote {\bibinfo {title} {Too much
  theory leads economists to bad predictions},}\ }\href@noop {} {\bibfield
  {journal} {\bibinfo  {journal} {Wired}\ } (\bibinfo {year} {2019})},\
  \bibinfo {note} {accessed 2019-05-20}\BibitemShut {NoStop}%
\bibitem [{\citenamefont {Popper}(1986)}]{popper1986the}%
  \BibitemOpen
  \bibfield  {author} {\bibinfo {author} {\bibfnamefont {Karl}\ \bibnamefont
  {Popper}},\ }\href@noop {} {\emph {\bibinfo {title} {The poverty of
  historicism}}}\ (\bibinfo  {publisher} {Ark},\ \bibinfo {address} {London},\
  \bibinfo {year} {1986})\BibitemShut {NoStop}%
\bibitem [{\citenamefont {Halpern}(2016)}]{halpern2016inside}%
  \BibitemOpen
  \bibfield  {author} {\bibinfo {author} {\bibfnamefont {David}\ \bibnamefont
  {Halpern}},\ }\href@noop {} {\emph {\bibinfo {title} {Inside the nudge unit :
  how small changes can make a big difference}}}\ (\bibinfo  {publisher} {WH
  Allen},\ \bibinfo {address} {London},\ \bibinfo {year} {2016})\BibitemShut
  {NoStop}%
\bibitem [{\citenamefont {Levitt}\ and\ \citenamefont
  {List}(2008)}]{levitt2008homo}%
  \BibitemOpen
  \bibfield  {author} {\bibinfo {author} {\bibfnamefont {Steven~D}\
  \bibnamefont {Levitt}}\ and\ \bibinfo {author} {\bibfnamefont {John~A}\
  \bibnamefont {List}},\ }\bibfield  {title} {\enquote {\bibinfo {title} {Homo
  economicus evolves},}\ }\href@noop {} {\bibfield  {journal} {\bibinfo
  {journal} {Science}\ }\textbf {\bibinfo {volume} {319}},\ \bibinfo {pages}
  {909--910} (\bibinfo {year} {2008})}\BibitemShut {NoStop}%
\bibitem [{\citenamefont {Thaler}(2015)}]{thaler2015misbehaving}%
  \BibitemOpen
  \bibfield  {author} {\bibinfo {author} {\bibfnamefont {Richard}\ \bibnamefont
  {Thaler}},\ }\href@noop {} {\emph {\bibinfo {title} {Misbehaving: The Making
  of Behavioral Economics}}},\ Business/Economics\ (\bibinfo  {publisher} {W.W.
  Norton},\ \bibinfo {year} {2015})\BibitemShut {NoStop}%
\bibitem [{\citenamefont {Simon}(1955)}]{simon1955behavioral}%
  \BibitemOpen
  \bibfield  {author} {\bibinfo {author} {\bibfnamefont {Herbert~A}\
  \bibnamefont {Simon}},\ }\bibfield  {title} {\enquote {\bibinfo {title} {A
  behavioral model of rational choice},}\ }\href@noop {} {\bibfield  {journal}
  {\bibinfo  {journal} {The Quarterly Journal of Economics}\ }\textbf {\bibinfo
  {volume} {69}},\ \bibinfo {pages} {99--118} (\bibinfo {year}
  {1955})}\BibitemShut {NoStop}%
\bibitem [{\citenamefont {Tversky}\ and\ \citenamefont
  {Kahneman}(1974)}]{Tversky1974}%
  \BibitemOpen
  \bibfield  {author} {\bibinfo {author} {\bibfnamefont {Amos}\ \bibnamefont
  {Tversky}}\ and\ \bibinfo {author} {\bibfnamefont {Daniel}\ \bibnamefont
  {Kahneman}},\ }\bibfield  {title} {\enquote {\bibinfo {title} {Judgment under
  uncertainty: Heuristics and biases},}\ }\href@noop {} {\bibfield  {journal}
  {\bibinfo  {journal} {Science}\ }\textbf {\bibinfo {volume} {185}},\ \bibinfo
  {pages} {1124--1131} (\bibinfo {year} {1974})}\BibitemShut {NoStop}%
\bibitem [{\citenamefont {Kahneman}(2011)}]{kahneman2011thinking}%
  \BibitemOpen
  \bibfield  {author} {\bibinfo {author} {\bibfnamefont {Daniel}\ \bibnamefont
  {Kahneman}},\ }\href@noop {} {\emph {\bibinfo {title} {Thinking, fast and
  slow}}}\ (\bibinfo  {publisher} {Farrar, Straus and Giroux},\ \bibinfo {year}
  {2011})\BibitemShut {NoStop}%
\bibitem [{\citenamefont {Wilkinson}\ and\ \citenamefont
  {Klaes}(2012)}]{wilkinson2012}%
  \BibitemOpen
  \bibfield  {author} {\bibinfo {author} {\bibfnamefont {Nick}\ \bibnamefont
  {Wilkinson}}\ and\ \bibinfo {author} {\bibfnamefont {Matthias}\ \bibnamefont
  {Klaes}},\ }\href@noop {} {\emph {\bibinfo {title} {An introduction to
  behavioral economics}}}\ (\bibinfo  {publisher} {Palgrave Macmillan},\
  \bibinfo {address} {Houndmills, Basingstoke New York, NY},\ \bibinfo {year}
  {2012})\BibitemShut {NoStop}%
\bibitem [{\citenamefont {Just}(2014)}]{just2014introduction}%
  \BibitemOpen
  \bibfield  {author} {\bibinfo {author} {\bibfnamefont {David}\ \bibnamefont
  {Just}},\ }\href@noop {} {\emph {\bibinfo {title} {Introduction to behavioral
  economics : noneconomic factors that shape economic decisions}}}\ (\bibinfo
  {publisher} {Wiley},\ \bibinfo {address} {Hoboken, NJ},\ \bibinfo {year}
  {2014})\BibitemShut {NoStop}%
\bibitem [{\citenamefont {Thaler}\ and\ \citenamefont
  {Sunstein}(2008)}]{thaler2008nudge}%
  \BibitemOpen
  \bibfield  {author} {\bibinfo {author} {\bibfnamefont {Richard}\ \bibnamefont
  {Thaler}}\ and\ \bibinfo {author} {\bibfnamefont {Cass}\ \bibnamefont
  {Sunstein}},\ }\href@noop {} {\emph {\bibinfo {title} {Nudge : improving
  decisions about health, wealth, and happiness}}}\ (\bibinfo  {publisher}
  {Yale University Press},\ \bibinfo {year} {2008})\BibitemShut {NoStop}%
\bibitem [{\citenamefont {Thaler}(1980)}]{thaler1980toward}%
  \BibitemOpen
  \bibfield  {author} {\bibinfo {author} {\bibfnamefont {Richard}\ \bibnamefont
  {Thaler}},\ }\bibfield  {title} {\enquote {\bibinfo {title} {Toward a
  positive theory of consumer choice},}\ }\href@noop {} {\bibfield  {journal}
  {\bibinfo  {journal} {Journal of Economic Behavior \& Organization}\ }\textbf
  {\bibinfo {volume} {1}},\ \bibinfo {pages} {39--60} (\bibinfo {year}
  {1980})}\BibitemShut {NoStop}%
\bibitem [{\citenamefont {Sunstein}(2017)}]{sunstein2017human}%
  \BibitemOpen
  \bibfield  {author} {\bibinfo {author} {\bibfnamefont {Cass}\ \bibnamefont
  {Sunstein}},\ }\href@noop {} {\emph {\bibinfo {title} {Human Agency and
  Behavioral Economics Nudging Fast and Slow}}}\ (\bibinfo  {publisher}
  {Palgrave Macmillan},\ \bibinfo {address} {Cham, Switzerland},\ \bibinfo
  {year} {2017})\BibitemShut {NoStop}%
\bibitem [{\citenamefont {Gino}(2017)}]{gino2017}%
  \BibitemOpen
  \bibfield  {author} {\bibinfo {author} {\bibfnamefont {Francesca}\
  \bibnamefont {Gino}},\ }\bibfield  {title} {\enquote {\bibinfo {title} {The
  rise of behavioral economics and its influence on organizations},}\
  }\href@noop {} {\bibfield  {journal} {\bibinfo  {journal} {Harvard Business
  Review}\ } (\bibinfo {year} {2017})},\ \bibinfo {note} {accessed
  2019-05-19}\BibitemShut {NoStop}%
\bibitem [{\citenamefont {Chetty}(2015)}]{Chetty2015}%
  \BibitemOpen
  \bibfield  {author} {\bibinfo {author} {\bibfnamefont {Raj}\ \bibnamefont
  {Chetty}},\ }\bibfield  {title} {\enquote {\bibinfo {title} {Behavioral
  economics and public policy: A pragmatic perspective},}\ }\href@noop {}
  {\bibfield  {journal} {\bibinfo  {journal} {American Economic Review}\
  }\textbf {\bibinfo {volume} {105}},\ \bibinfo {pages} {1--33} (\bibinfo
  {year} {2015})}\BibitemShut {NoStop}%
\bibitem [{\citenamefont {Congdon}\ \emph {et~al.}(2011)\citenamefont
  {Congdon}, \citenamefont {Kling},\ and\ \citenamefont
  {Mullainathan}}]{congdon2011policy}%
  \BibitemOpen
  \bibfield  {author} {\bibinfo {author} {\bibfnamefont {William}\ \bibnamefont
  {Congdon}}, \bibinfo {author} {\bibfnamefont {Jeffrey~R}\ \bibnamefont
  {Kling}}, \ and\ \bibinfo {author} {\bibfnamefont {Sendhil}\ \bibnamefont
  {Mullainathan}},\ }\href@noop {} {\emph {\bibinfo {title} {Policy and choice
  : public finance through the lens of behavioral economics}}}\ (\bibinfo
  {publisher} {Brookings Institution Press},\ \bibinfo {address} {Washington,
  D.C},\ \bibinfo {year} {2011})\BibitemShut {NoStop}%
\bibitem [{\citenamefont {Madrian}(2014)}]{madrian2014}%
  \BibitemOpen
  \bibfield  {author} {\bibinfo {author} {\bibfnamefont {Brigitte~C}\
  \bibnamefont {Madrian}},\ }\bibfield  {title} {\enquote {\bibinfo {title}
  {Applying insights from behavioral economics to policy design},}\ }\href@noop
  {} {\bibfield  {journal} {\bibinfo  {journal} {Annual Review of Economics}\
  }\textbf {\bibinfo {volume} {6}},\ \bibinfo {pages} {663--688} (\bibinfo
  {year} {2014})}\BibitemShut {NoStop}%
\bibitem [{\citenamefont {Sunstein}(2016)}]{sunstein2016}%
  \BibitemOpen
  \bibfield  {author} {\bibinfo {author} {\bibfnamefont {Cass}\ \bibnamefont
  {Sunstein}},\ }\href@noop {} {\emph {\bibinfo {title} {The ethics of
  influence : government in the age of behavioral science}}}\ (\bibinfo
  {publisher} {Cambridge University Press},\ \bibinfo {address} {New York, NY,
  USA},\ \bibinfo {year} {2016})\BibitemShut {NoStop}%
\bibitem [{\citenamefont {Guszcza}(2015)}]{guszcza2015}%
  \BibitemOpen
  \bibfield  {author} {\bibinfo {author} {\bibfnamefont {Jim}\ \bibnamefont
  {Guszcza}},\ }\bibfield  {title} {\enquote {\bibinfo {title} {The last-mile
  problem: how data science and behavioral science can work together},}\
  }\href@noop {} {\bibfield  {journal} {\bibinfo  {journal} {Deloitte Review}\
  } (\bibinfo {year} {2015})},\ \bibinfo {note} {accessed
  2019-05-19}\BibitemShut {NoStop}%
\bibitem [{\citenamefont {Davenport}\ \emph {et~al.}(2018)\citenamefont
  {Davenport}, \citenamefont {Guszcza},\ and\ \citenamefont
  {Szwartz}}]{davenport2018}%
  \BibitemOpen
  \bibfield  {author} {\bibinfo {author} {\bibfnamefont {Thomas~H}\
  \bibnamefont {Davenport}}, \bibinfo {author} {\bibfnamefont {James}\
  \bibnamefont {Guszcza}}, \ and\ \bibinfo {author} {\bibfnamefont {Greg}\
  \bibnamefont {Szwartz}},\ }\bibfield  {title} {\enquote {\bibinfo {title}
  {Using behavioral nudges to treat diabetes},}\ }\href@noop {} {\bibfield
  {journal} {\bibinfo  {journal} {Harvard Business Review}\ } (\bibinfo {year}
  {2018})},\ \bibinfo {note} {accessed 2019-05-19}\BibitemShut {NoStop}%
\bibitem [{\citenamefont {Benartzi}\ \emph {et~al.}(2017)\citenamefont
  {Benartzi}, \citenamefont {Beshears}, \citenamefont {Milkman}, \citenamefont
  {Sunstein}, \citenamefont {Thaler}, \citenamefont {Shankar}, \citenamefont
  {Tucker-Ray}, \citenamefont {Congdon},\ and\ \citenamefont
  {Galing}}]{benartzi2017}%
  \BibitemOpen
  \bibfield  {author} {\bibinfo {author} {\bibfnamefont {Shlomo}\ \bibnamefont
  {Benartzi}}, \bibinfo {author} {\bibfnamefont {John}\ \bibnamefont
  {Beshears}}, \bibinfo {author} {\bibfnamefont {Katherine~L}\ \bibnamefont
  {Milkman}}, \bibinfo {author} {\bibfnamefont {Cass~R}\ \bibnamefont
  {Sunstein}}, \bibinfo {author} {\bibfnamefont {Richard~H}\ \bibnamefont
  {Thaler}}, \bibinfo {author} {\bibfnamefont {Maya}\ \bibnamefont {Shankar}},
  \bibinfo {author} {\bibfnamefont {Will}\ \bibnamefont {Tucker-Ray}}, \bibinfo
  {author} {\bibfnamefont {William~J}\ \bibnamefont {Congdon}}, \ and\ \bibinfo
  {author} {\bibfnamefont {Steven}\ \bibnamefont {Galing}},\ }\bibfield
  {title} {\enquote {\bibinfo {title} {Should governments invest more in
  nudging?}}\ }\href@noop {} {\bibfield  {journal} {\bibinfo  {journal}
  {Psychological science}\ }\textbf {\bibinfo {volume} {28}},\ \bibinfo {pages}
  {1041--1055} (\bibinfo {year} {2017})}\BibitemShut {NoStop}%
\bibitem [{\citenamefont {Russell}\ and\ \citenamefont
  {Norvig}(2016)}]{russell2016}%
  \BibitemOpen
  \bibfield  {author} {\bibinfo {author} {\bibfnamefont {Stuart}\ \bibnamefont
  {Russell}}\ and\ \bibinfo {author} {\bibfnamefont {Peter}\ \bibnamefont
  {Norvig}},\ }\href@noop {} {\emph {\bibinfo {title} {Artificial intelligence:
  a modern approach}}}\ (\bibinfo  {publisher} {Pearson; 3 edition},\ \bibinfo
  {year} {2016})\BibitemShut {NoStop}%
\bibitem [{\citenamefont {Parkes}\ and\ \citenamefont
  {Wellman}(2015)}]{parkes2015}%
  \BibitemOpen
  \bibfield  {author} {\bibinfo {author} {\bibfnamefont {David~C}\ \bibnamefont
  {Parkes}}\ and\ \bibinfo {author} {\bibfnamefont {Michael~P}\ \bibnamefont
  {Wellman}},\ }\bibfield  {title} {\enquote {\bibinfo {title} {Economic
  reasoning and artificial intelligence},}\ }\href@noop {} {\bibfield
  {journal} {\bibinfo  {journal} {Science}\ }\textbf {\bibinfo {volume}
  {349}},\ \bibinfo {pages} {267--272} (\bibinfo {year} {2015})}\BibitemShut
  {NoStop}%
\bibitem [{\citenamefont {Goodfellow}(2016)}]{goodfellow2016deep}%
  \BibitemOpen
  \bibfield  {author} {\bibinfo {author} {\bibfnamefont {Ian}\ \bibnamefont
  {Goodfellow}},\ }\href@noop {} {\emph {\bibinfo {title} {Deep learning}}}\
  (\bibinfo  {publisher} {The MIT Press},\ \bibinfo {address} {Cambridge,
  Massachusetts},\ \bibinfo {year} {2016})\BibitemShut {NoStop}%
\bibitem [{\citenamefont {Hastie}\ \emph {et~al.}(2009)\citenamefont {Hastie},
  \citenamefont {Tibshirani},\ and\ \citenamefont {Friedman}}]{hastie2009the}%
  \BibitemOpen
  \bibfield  {author} {\bibinfo {author} {\bibfnamefont {Trevor}\ \bibnamefont
  {Hastie}}, \bibinfo {author} {\bibfnamefont {Robert}\ \bibnamefont
  {Tibshirani}}, \ and\ \bibinfo {author} {\bibfnamefont {Jerome}\ \bibnamefont
  {Friedman}},\ }\href@noop {} {\emph {\bibinfo {title} {The elements of
  statistical learning : Data mining, inference, and prediction}}}\ (\bibinfo
  {publisher} {Springer},\ \bibinfo {address} {New York},\ \bibinfo {year}
  {2009})\BibitemShut {NoStop}%
\bibitem [{\citenamefont {Nilsson}(2010)}]{nilsson2010the}%
  \BibitemOpen
  \bibfield  {author} {\bibinfo {author} {\bibfnamefont {Nils}\ \bibnamefont
  {Nilsson}},\ }\href@noop {} {\emph {\bibinfo {title} {The quest for
  artificial intelligence : a history of ideas and achievements}}}\ (\bibinfo
  {publisher} {Cambridge University Press},\ \bibinfo {address} {Cambridge New
  York},\ \bibinfo {year} {2010})\BibitemShut {NoStop}%
\bibitem [{\citenamefont {Simon}(1996)}]{simon1996the}%
  \BibitemOpen
  \bibfield  {author} {\bibinfo {author} {\bibfnamefont {Herbert}\ \bibnamefont
  {Simon}},\ }\href@noop {} {\emph {\bibinfo {title} {The sciences of the
  artificial, 3rd ed.}}}\ (\bibinfo  {publisher} {MIT Press},\ \bibinfo
  {address} {Cambridge, Mass},\ \bibinfo {year} {1996})\BibitemShut {NoStop}%
\bibitem [{\citenamefont {Einav}\ and\ \citenamefont
  {Levin}(2014)}]{einav2014}%
  \BibitemOpen
  \bibfield  {author} {\bibinfo {author} {\bibfnamefont {Liran}\ \bibnamefont
  {Einav}}\ and\ \bibinfo {author} {\bibfnamefont {Jonathan}\ \bibnamefont
  {Levin}},\ }\bibfield  {title} {\enquote {\bibinfo {title} {Economics in the
  age of big data},}\ }\href@noop {} {\bibfield  {journal} {\bibinfo  {journal}
  {Science}\ }\textbf {\bibinfo {volume} {346}},\ \bibinfo {pages} {1243089}
  (\bibinfo {year} {2014})}\BibitemShut {NoStop}%
\bibitem [{\citenamefont {Provost}\ and\ \citenamefont
  {Fawcett}(2013)}]{provost2013data}%
  \BibitemOpen
  \bibfield  {author} {\bibinfo {author} {\bibfnamefont {Foster}\ \bibnamefont
  {Provost}}\ and\ \bibinfo {author} {\bibfnamefont {Tom}\ \bibnamefont
  {Fawcett}},\ }\href@noop {} {\emph {\bibinfo {title} {Data science for
  business : what you need to know about data mining and data-analytic
  thinking}}}\ (\bibinfo  {publisher} {O'Reilly},\ \bibinfo {address}
  {Sebastopol, Calif},\ \bibinfo {year} {2013})\BibitemShut {NoStop}%
\bibitem [{\citenamefont {Jordan}\ and\ \citenamefont
  {Mitchell}(2015)}]{jordan2015}%
  \BibitemOpen
  \bibfield  {author} {\bibinfo {author} {\bibfnamefont {M.~I.}\ \bibnamefont
  {Jordan}}\ and\ \bibinfo {author} {\bibfnamefont {T.~M.}\ \bibnamefont
  {Mitchell}},\ }\bibfield  {title} {\enquote {\bibinfo {title} {Machine
  learning: Trends, perspectives, and prospects},}\ }\href@noop {} {\bibfield
  {journal} {\bibinfo  {journal} {Science}\ }\textbf {\bibinfo {volume}
  {349}},\ \bibinfo {pages} {255--260} (\bibinfo {year} {2015})}\BibitemShut
  {NoStop}%
\bibitem [{\citenamefont {LeCun}\ \emph {et~al.}(2015)\citenamefont {LeCun},
  \citenamefont {Bengio},\ and\ \citenamefont {Hinton}}]{lecun2015}%
  \BibitemOpen
  \bibfield  {author} {\bibinfo {author} {\bibfnamefont {Yann}\ \bibnamefont
  {LeCun}}, \bibinfo {author} {\bibfnamefont {Yoshua}\ \bibnamefont {Bengio}},
  \ and\ \bibinfo {author} {\bibfnamefont {Geoffrey}\ \bibnamefont {Hinton}},\
  }\bibfield  {title} {\enquote {\bibinfo {title} {Deep learning},}\
  }\href@noop {} {\bibfield  {journal} {\bibinfo  {journal} {Nature}\ }\textbf
  {\bibinfo {volume} {521}},\ \bibinfo {pages} {436} (\bibinfo {year}
  {2015})}\BibitemShut {NoStop}%
\bibitem [{\citenamefont {Grimmer}\ and\ \citenamefont
  {Stewart}(2013)}]{grimmer2013}%
  \BibitemOpen
  \bibfield  {author} {\bibinfo {author} {\bibfnamefont {Justin}\ \bibnamefont
  {Grimmer}}\ and\ \bibinfo {author} {\bibfnamefont {Brandon~M}\ \bibnamefont
  {Stewart}},\ }\bibfield  {title} {\enquote {\bibinfo {title} {Text as data:
  The promise and pitfalls of automatic content analysis methods for political
  texts},}\ }\href@noop {} {\bibfield  {journal} {\bibinfo  {journal}
  {Political analysis}\ }\textbf {\bibinfo {volume} {21}},\ \bibinfo {pages}
  {267--297} (\bibinfo {year} {2013})}\BibitemShut {NoStop}%
\bibitem [{\citenamefont {{Wang}}\ \emph {et~al.}(2016)\citenamefont {{Wang}},
  \citenamefont {{Khosla}}, \citenamefont {{Gargeya}}, \citenamefont
  {{Irshad}},\ and\ \citenamefont {{Beck}}}]{wang2016}%
  \BibitemOpen
  \bibfield  {author} {\bibinfo {author} {\bibfnamefont {Dayong}\ \bibnamefont
  {{Wang}}}, \bibinfo {author} {\bibfnamefont {Aditya}\ \bibnamefont
  {{Khosla}}}, \bibinfo {author} {\bibfnamefont {Rishab}\ \bibnamefont
  {{Gargeya}}}, \bibinfo {author} {\bibfnamefont {Humayun}\ \bibnamefont
  {{Irshad}}}, \ and\ \bibinfo {author} {\bibfnamefont {Andrew~H.}\
  \bibnamefont {{Beck}}},\ }\bibfield  {title} {\enquote {\bibinfo {title}
  {{Deep Learning for Identifying Metastatic Breast Cancer}},}\ }\href@noop {}
  {\bibfield  {journal} {\bibinfo  {journal} {arXiv e-prints}\ ,\ \bibinfo
  {eid} {arXiv:1606.05718}} (\bibinfo {year} {2016})}\BibitemShut {NoStop}%
\bibitem [{\citenamefont {Bajari}\ \emph {et~al.}(2015)\citenamefont {Bajari},
  \citenamefont {Nekipelov}, \citenamefont {Ryan},\ and\ \citenamefont
  {Yang}}]{bajari2015machine}%
  \BibitemOpen
  \bibfield  {author} {\bibinfo {author} {\bibfnamefont {Patrick}\ \bibnamefont
  {Bajari}}, \bibinfo {author} {\bibfnamefont {Denis}\ \bibnamefont
  {Nekipelov}}, \bibinfo {author} {\bibfnamefont {Stephen~P}\ \bibnamefont
  {Ryan}}, \ and\ \bibinfo {author} {\bibfnamefont {Miaoyu}\ \bibnamefont
  {Yang}},\ }\bibfield  {title} {\enquote {\bibinfo {title} {Machine learning
  methods for demand estimation},}\ }\href@noop {} {\bibfield  {journal}
  {\bibinfo  {journal} {American Economic Review}\ }\textbf {\bibinfo {volume}
  {105}},\ \bibinfo {pages} {481--85} (\bibinfo {year} {2015})}\BibitemShut
  {NoStop}%
\bibitem [{\citenamefont {Varian}(2014)}]{Varian2014}%
  \BibitemOpen
  \bibfield  {author} {\bibinfo {author} {\bibfnamefont {Hal~R}\ \bibnamefont
  {Varian}},\ }\bibfield  {title} {\enquote {\bibinfo {title} {Big data: New
  tricks for econometrics},}\ }\href@noop {} {\bibfield  {journal} {\bibinfo
  {journal} {Journal of Economic Perspectives}\ }\textbf {\bibinfo {volume}
  {28}},\ \bibinfo {pages} {3 --28} (\bibinfo {year} {2014})}\BibitemShut
  {NoStop}%
\bibitem [{\citenamefont {Pariser}(2011)}]{pariser2011}%
  \BibitemOpen
  \bibfield  {author} {\bibinfo {author} {\bibfnamefont {Eli}\ \bibnamefont
  {Pariser}},\ }\bibfield  {title} {\enquote {\bibinfo {title} {Welcome to the
  brave new world of persuasion profiling},}\ }\href@noop {} {\bibfield
  {journal} {\bibinfo  {journal} {Wired}\ } (\bibinfo {year} {2011})},\
  \bibinfo {note} {accessed 2019-05-03}\BibitemShut {NoStop}%
\bibitem [{\citenamefont {Dressel}\ and\ \citenamefont
  {Farid}(2018)}]{Dressel2018}%
  \BibitemOpen
  \bibfield  {author} {\bibinfo {author} {\bibfnamefont {Julia}\ \bibnamefont
  {Dressel}}\ and\ \bibinfo {author} {\bibfnamefont {Hany}\ \bibnamefont
  {Farid}},\ }\bibfield  {title} {\enquote {\bibinfo {title} {The accuracy,
  fairness, and limits of predicting recidivism},}\ }\href@noop {} {\bibfield
  {journal} {\bibinfo  {journal} {Science Advances}\ }\textbf {\bibinfo
  {volume} {4}} (\bibinfo {year} {2018})}\BibitemShut {NoStop}%
\bibitem [{\citenamefont {Chandler}\ \emph {et~al.}(2011)\citenamefont
  {Chandler}, \citenamefont {Levitt},\ and\ \citenamefont
  {List}}]{chandler2011predicting}%
  \BibitemOpen
  \bibfield  {author} {\bibinfo {author} {\bibfnamefont {Dana}\ \bibnamefont
  {Chandler}}, \bibinfo {author} {\bibfnamefont {Steven~D}\ \bibnamefont
  {Levitt}}, \ and\ \bibinfo {author} {\bibfnamefont {John~A}\ \bibnamefont
  {List}},\ }\bibfield  {title} {\enquote {\bibinfo {title} {Predicting and
  preventing shootings among at-risk youth},}\ }\href@noop {} {\bibfield
  {journal} {\bibinfo  {journal} {American Economic Review}\ }\textbf {\bibinfo
  {volume} {101}},\ \bibinfo {pages} {288--92} (\bibinfo {year}
  {2011})}\BibitemShut {NoStop}%
\bibitem [{\citenamefont {Blumenstock}\ \emph {et~al.}(2015)\citenamefont
  {Blumenstock}, \citenamefont {Cadamuro},\ and\ \citenamefont
  {On}}]{blumenstock2015predicting}%
  \BibitemOpen
  \bibfield  {author} {\bibinfo {author} {\bibfnamefont {Joshua}\ \bibnamefont
  {Blumenstock}}, \bibinfo {author} {\bibfnamefont {Gabriel}\ \bibnamefont
  {Cadamuro}}, \ and\ \bibinfo {author} {\bibfnamefont {Robert}\ \bibnamefont
  {On}},\ }\bibfield  {title} {\enquote {\bibinfo {title} {Predicting poverty
  and wealth from mobile phone metadata},}\ }\href@noop {} {\bibfield
  {journal} {\bibinfo  {journal} {Science}\ }\textbf {\bibinfo {volume}
  {350}},\ \bibinfo {pages} {1073--1076} (\bibinfo {year} {2015})}\BibitemShut
  {NoStop}%
\bibitem [{\citenamefont {Blumenstock}(2016)}]{blumenstock2016}%
  \BibitemOpen
  \bibfield  {author} {\bibinfo {author} {\bibfnamefont {Joshua}\ \bibnamefont
  {Blumenstock}},\ }\bibfield  {title} {\enquote {\bibinfo {title} {Fighting
  poverty with data},}\ }\href@noop {} {\bibfield  {journal} {\bibinfo
  {journal} {Science}\ }\textbf {\bibinfo {volume} {353}},\ \bibinfo {pages}
  {753 -- 754} (\bibinfo {year} {2016})}\BibitemShut {NoStop}%
\bibitem [{\citenamefont {Jean}\ \emph {et~al.}(2016)\citenamefont {Jean},
  \citenamefont {Burke}, \citenamefont {Xie}, \citenamefont {Davis},
  \citenamefont {Lobell},\ and\ \citenamefont {Ermon}}]{Jean2016}%
  \BibitemOpen
  \bibfield  {author} {\bibinfo {author} {\bibfnamefont {Neal}\ \bibnamefont
  {Jean}}, \bibinfo {author} {\bibfnamefont {Marshall}\ \bibnamefont {Burke}},
  \bibinfo {author} {\bibfnamefont {Michael}\ \bibnamefont {Xie}}, \bibinfo
  {author} {\bibfnamefont {W~Matthew}\ \bibnamefont {Davis}}, \bibinfo {author}
  {\bibfnamefont {David~B}\ \bibnamefont {Lobell}}, \ and\ \bibinfo {author}
  {\bibfnamefont {Stefano}\ \bibnamefont {Ermon}},\ }\bibfield  {title}
  {\enquote {\bibinfo {title} {Combining satellite imagery and machine learning
  to predict poverty},}\ }\href@noop {} {\bibfield  {journal} {\bibinfo
  {journal} {Science}\ }\textbf {\bibinfo {volume} {353}},\ \bibinfo {pages}
  {790--794} (\bibinfo {year} {2016})}\BibitemShut {NoStop}%
\bibitem [{\citenamefont {Brynjolfsson}\ and\ \citenamefont
  {Mitchell}(2017)}]{Brynjolfsson2017}%
  \BibitemOpen
  \bibfield  {author} {\bibinfo {author} {\bibfnamefont {Erik}\ \bibnamefont
  {Brynjolfsson}}\ and\ \bibinfo {author} {\bibfnamefont {Tom}\ \bibnamefont
  {Mitchell}},\ }\bibfield  {title} {\enquote {\bibinfo {title} {{What can
  machine learning do? Workforce implications}},}\ }\href@noop {} {\bibfield
  {journal} {\bibinfo  {journal} {Science}\ }\textbf {\bibinfo {volume}
  {358}},\ \bibinfo {pages} {1530--1534} (\bibinfo {year} {2017})}\BibitemShut
  {NoStop}%
\bibitem [{\citenamefont {Bonnefon}\ \emph {et~al.}(2016)\citenamefont
  {Bonnefon}, \citenamefont {Shariff},\ and\ \citenamefont
  {Rahwan}}]{bonnefon2016social}%
  \BibitemOpen
  \bibfield  {author} {\bibinfo {author} {\bibfnamefont {Jean-Fran{\c{c}}ois}\
  \bibnamefont {Bonnefon}}, \bibinfo {author} {\bibfnamefont {Azim}\
  \bibnamefont {Shariff}}, \ and\ \bibinfo {author} {\bibfnamefont {Iyad}\
  \bibnamefont {Rahwan}},\ }\bibfield  {title} {\enquote {\bibinfo {title} {The
  social dilemma of autonomous vehicles},}\ }\href@noop {} {\bibfield
  {journal} {\bibinfo  {journal} {Science}\ }\textbf {\bibinfo {volume}
  {352}},\ \bibinfo {pages} {1573--1576} (\bibinfo {year} {2016})}\BibitemShut
  {NoStop}%
\bibitem [{\citenamefont {Gerrish}(2018)}]{gerrish2018how}%
  \BibitemOpen
  \bibfield  {author} {\bibinfo {author} {\bibfnamefont {Sean}\ \bibnamefont
  {Gerrish}},\ }\href@noop {} {\emph {\bibinfo {title} {How smart machines
  think}}}\ (\bibinfo  {publisher} {The MIT Press},\ \bibinfo {address}
  {Cambridge, MA},\ \bibinfo {year} {2018})\BibitemShut {NoStop}%
\bibitem [{\citenamefont {Huang}\ \emph {et~al.}(2007)\citenamefont {Huang},
  \citenamefont {Chen},\ and\ \citenamefont {Wang}}]{huang2007credit}%
  \BibitemOpen
  \bibfield  {author} {\bibinfo {author} {\bibfnamefont {Cheng-Lung}\
  \bibnamefont {Huang}}, \bibinfo {author} {\bibfnamefont {Mu-Chen}\
  \bibnamefont {Chen}}, \ and\ \bibinfo {author} {\bibfnamefont {Chieh-Jen}\
  \bibnamefont {Wang}},\ }\bibfield  {title} {\enquote {\bibinfo {title}
  {Credit scoring with a data mining approach based on support vector
  machines},}\ }\href@noop {} {\bibfield  {journal} {\bibinfo  {journal}
  {Expert Systems with Applications}\ }\textbf {\bibinfo {volume} {33}},\
  \bibinfo {pages} {847--856} (\bibinfo {year} {2007})}\BibitemShut {NoStop}%
\bibitem [{\citenamefont {Wellman}\ and\ \citenamefont
  {Rajan}(2017)}]{wellman2017ethical}%
  \BibitemOpen
  \bibfield  {author} {\bibinfo {author} {\bibfnamefont {Michael~P}\
  \bibnamefont {Wellman}}\ and\ \bibinfo {author} {\bibfnamefont {Uday}\
  \bibnamefont {Rajan}},\ }\bibfield  {title} {\enquote {\bibinfo {title}
  {Ethical issues for autonomous trading agents},}\ }\href@noop {} {\bibfield
  {journal} {\bibinfo  {journal} {Minds and Machines}\ }\textbf {\bibinfo
  {volume} {27}},\ \bibinfo {pages} {609--624} (\bibinfo {year}
  {2017})}\BibitemShut {NoStop}%
\bibitem [{\citenamefont {Deng}\ \emph {et~al.}(2016)\citenamefont {Deng},
  \citenamefont {Bao}, \citenamefont {Kong}, \citenamefont {Ren},\ and\
  \citenamefont {Dai}}]{deng2016deep}%
  \BibitemOpen
  \bibfield  {author} {\bibinfo {author} {\bibfnamefont {Yue}\ \bibnamefont
  {Deng}}, \bibinfo {author} {\bibfnamefont {Feng}\ \bibnamefont {Bao}},
  \bibinfo {author} {\bibfnamefont {Youyong}\ \bibnamefont {Kong}}, \bibinfo
  {author} {\bibfnamefont {Zhiquan}\ \bibnamefont {Ren}}, \ and\ \bibinfo
  {author} {\bibfnamefont {Qionghai}\ \bibnamefont {Dai}},\ }\bibfield  {title}
  {\enquote {\bibinfo {title} {Deep direct reinforcement learning for financial
  signal representation and trading},}\ }\href@noop {} {\bibfield  {journal}
  {\bibinfo  {journal} {IEEE Transactions on Neural Networks and Learning
  Systems}\ }\textbf {\bibinfo {volume} {28}},\ \bibinfo {pages} {653--664}
  (\bibinfo {year} {2016})}\BibitemShut {NoStop}%
\bibitem [{\citenamefont {Metz}(2016)}]{metz2016}%
  \BibitemOpen
  \bibfield  {author} {\bibinfo {author} {\bibfnamefont {Cade}\ \bibnamefont
  {Metz}},\ }\bibfield  {title} {\enquote {\bibinfo {title} {The rise of the
  artificially intelligent hedge fund},}\ }\href@noop {} {\bibfield  {journal}
  {\bibinfo  {journal} {Wired}\ } (\bibinfo {year} {2016})},\ \bibinfo {note}
  {accessed 2019-05-03}\BibitemShut {NoStop}%
\bibitem [{\citenamefont {Heaton}\ \emph {et~al.}(2017)\citenamefont {Heaton},
  \citenamefont {Polson},\ and\ \citenamefont {Witte}}]{heaton2017deep}%
  \BibitemOpen
  \bibfield  {author} {\bibinfo {author} {\bibfnamefont {JB}~\bibnamefont
  {Heaton}}, \bibinfo {author} {\bibfnamefont {NG}~\bibnamefont {Polson}}, \
  and\ \bibinfo {author} {\bibfnamefont {Jan~Hendrik}\ \bibnamefont {Witte}},\
  }\bibfield  {title} {\enquote {\bibinfo {title} {Deep learning for finance:
  deep portfolios},}\ }\href@noop {} {\bibfield  {journal} {\bibinfo  {journal}
  {Applied Stochastic Models in Business and Industry}\ }\textbf {\bibinfo
  {volume} {33}},\ \bibinfo {pages} {3--12} (\bibinfo {year}
  {2017})}\BibitemShut {NoStop}%
\bibitem [{\citenamefont {Mullainathan}\ and\ \citenamefont
  {Spiess}(2017)}]{mullainathan2017machine}%
  \BibitemOpen
  \bibfield  {author} {\bibinfo {author} {\bibfnamefont {Sendhil}\ \bibnamefont
  {Mullainathan}}\ and\ \bibinfo {author} {\bibfnamefont {Jann}\ \bibnamefont
  {Spiess}},\ }\bibfield  {title} {\enquote {\bibinfo {title} {Machine
  learning: an applied econometric approach},}\ }\href@noop {} {\bibfield
  {journal} {\bibinfo  {journal} {Journal of Economic Perspectives}\ }\textbf
  {\bibinfo {volume} {31}},\ \bibinfo {pages} {87--106} (\bibinfo {year}
  {2017})}\BibitemShut {NoStop}%
\bibitem [{\citenamefont {Joel}\ \emph {et~al.}(2017)\citenamefont {Joel},
  \citenamefont {Eastwick},\ and\ \citenamefont {Finkel}}]{joel2017romantic}%
  \BibitemOpen
  \bibfield  {author} {\bibinfo {author} {\bibfnamefont {Samantha}\
  \bibnamefont {Joel}}, \bibinfo {author} {\bibfnamefont {Paul~W}\ \bibnamefont
  {Eastwick}}, \ and\ \bibinfo {author} {\bibfnamefont {Eli~J}\ \bibnamefont
  {Finkel}},\ }\bibfield  {title} {\enquote {\bibinfo {title} {Is romantic
  desire predictable? {M}achine learning applied to initial romantic
  attraction},}\ }\href@noop {} {\bibfield  {journal} {\bibinfo  {journal}
  {Psychological science}\ }\textbf {\bibinfo {volume} {28}},\ \bibinfo {pages}
  {1478--1489} (\bibinfo {year} {2017})}\BibitemShut {NoStop}%
\bibitem [{\citenamefont {Carrasquilla}\ and\ \citenamefont
  {Melko}(2017)}]{carrasquilla2017machine}%
  \BibitemOpen
  \bibfield  {author} {\bibinfo {author} {\bibfnamefont {Juan}\ \bibnamefont
  {Carrasquilla}}\ and\ \bibinfo {author} {\bibfnamefont {Roger~G}\
  \bibnamefont {Melko}},\ }\bibfield  {title} {\enquote {\bibinfo {title}
  {Machine learning phases of matter},}\ }\href@noop {} {\bibfield  {journal}
  {\bibinfo  {journal} {Nature Physics}\ }\textbf {\bibinfo {volume} {13}},\
  \bibinfo {pages} {431} (\bibinfo {year} {2017})}\BibitemShut {NoStop}%
\bibitem [{\citenamefont {Butler}\ \emph {et~al.}(2018)\citenamefont {Butler},
  \citenamefont {Davies}, \citenamefont {Cartwright}, \citenamefont {Isayev},\
  and\ \citenamefont {Walsh}}]{butler2018}%
  \BibitemOpen
  \bibfield  {author} {\bibinfo {author} {\bibfnamefont {Keith~T}\ \bibnamefont
  {Butler}}, \bibinfo {author} {\bibfnamefont {Daniel~W}\ \bibnamefont
  {Davies}}, \bibinfo {author} {\bibfnamefont {Hugh}\ \bibnamefont
  {Cartwright}}, \bibinfo {author} {\bibfnamefont {Olexandr}\ \bibnamefont
  {Isayev}}, \ and\ \bibinfo {author} {\bibfnamefont {Aron}\ \bibnamefont
  {Walsh}},\ }\bibfield  {title} {\enquote {\bibinfo {title} {Machine learning
  for molecular and materials science},}\ }\href@noop {} {\bibfield  {journal}
  {\bibinfo  {journal} {Nature}\ }\textbf {\bibinfo {volume} {559}},\ \bibinfo
  {pages} {547} (\bibinfo {year} {2018})}\BibitemShut {NoStop}%
\bibitem [{\citenamefont {Lookman}\ \emph {et~al.}(2019)\citenamefont
  {Lookman}, \citenamefont {Balachandran}, \citenamefont {Xue},\ and\
  \citenamefont {Yuan}}]{lookman2019active}%
  \BibitemOpen
  \bibfield  {author} {\bibinfo {author} {\bibfnamefont {Turab}\ \bibnamefont
  {Lookman}}, \bibinfo {author} {\bibfnamefont {Prasanna~V}\ \bibnamefont
  {Balachandran}}, \bibinfo {author} {\bibfnamefont {Dezhen}\ \bibnamefont
  {Xue}}, \ and\ \bibinfo {author} {\bibfnamefont {Ruihao}\ \bibnamefont
  {Yuan}},\ }\bibfield  {title} {\enquote {\bibinfo {title} {Active learning in
  materials science with emphasis on adaptive sampling using uncertainties for
  targeted design},}\ }\href@noop {} {\bibfield  {journal} {\bibinfo  {journal}
  {npj Computational Materials}\ }\textbf {\bibinfo {volume} {5}},\ \bibinfo
  {pages} {21} (\bibinfo {year} {2019})}\BibitemShut {NoStop}%
\bibitem [{\citenamefont {Biamonte}\ \emph {et~al.}(2017)\citenamefont
  {Biamonte}, \citenamefont {Wittek}, \citenamefont {Pancotti}, \citenamefont
  {Rebentrost}, \citenamefont {Wiebe},\ and\ \citenamefont
  {Lloyd}}]{biamonte2017quantum}%
  \BibitemOpen
  \bibfield  {author} {\bibinfo {author} {\bibfnamefont {Jacob}\ \bibnamefont
  {Biamonte}}, \bibinfo {author} {\bibfnamefont {Peter}\ \bibnamefont
  {Wittek}}, \bibinfo {author} {\bibfnamefont {Nicola}\ \bibnamefont
  {Pancotti}}, \bibinfo {author} {\bibfnamefont {Patrick}\ \bibnamefont
  {Rebentrost}}, \bibinfo {author} {\bibfnamefont {Nathan}\ \bibnamefont
  {Wiebe}}, \ and\ \bibinfo {author} {\bibfnamefont {Seth}\ \bibnamefont
  {Lloyd}},\ }\bibfield  {title} {\enquote {\bibinfo {title} {Quantum machine
  learning},}\ }\href@noop {} {\bibfield  {journal} {\bibinfo  {journal}
  {Nature}\ }\textbf {\bibinfo {volume} {549}},\ \bibinfo {pages} {195}
  (\bibinfo {year} {2017})}\BibitemShut {NoStop}%
\bibitem [{\citenamefont {Ramprasad}\ \emph {et~al.}(2017)\citenamefont
  {Ramprasad}, \citenamefont {Batra}, \citenamefont {Pilania}, \citenamefont
  {Mannodi-Kanakkithodi},\ and\ \citenamefont {Kim}}]{Ramprasad2017}%
  \BibitemOpen
  \bibfield  {author} {\bibinfo {author} {\bibfnamefont {Rampi}\ \bibnamefont
  {Ramprasad}}, \bibinfo {author} {\bibfnamefont {Rohit}\ \bibnamefont
  {Batra}}, \bibinfo {author} {\bibfnamefont {Ghanshyam}\ \bibnamefont
  {Pilania}}, \bibinfo {author} {\bibfnamefont {Arun}\ \bibnamefont
  {Mannodi-Kanakkithodi}}, \ and\ \bibinfo {author} {\bibfnamefont {Chiho}\
  \bibnamefont {Kim}},\ }\bibfield  {title} {\enquote {\bibinfo {title}
  {Machine learning in materials informatics: recent applications and
  prospects},}\ }\href@noop {} {\bibfield  {journal} {\bibinfo  {journal} {npj
  Computational Materials}\ }\textbf {\bibinfo {volume} {3}},\ \bibinfo {pages}
  {54} (\bibinfo {year} {2017})}\BibitemShut {NoStop}%
\bibitem [{\citenamefont {Oganov}\ \emph {et~al.}(2019)\citenamefont {Oganov},
  \citenamefont {Pickard}, \citenamefont {Zhu},\ and\ \citenamefont
  {Needs}}]{oganov2019structure}%
  \BibitemOpen
  \bibfield  {author} {\bibinfo {author} {\bibfnamefont {Artem~R}\ \bibnamefont
  {Oganov}}, \bibinfo {author} {\bibfnamefont {Chris~J}\ \bibnamefont
  {Pickard}}, \bibinfo {author} {\bibfnamefont {Qiang}\ \bibnamefont {Zhu}}, \
  and\ \bibinfo {author} {\bibfnamefont {Richard~J}\ \bibnamefont {Needs}},\
  }\bibfield  {title} {\enquote {\bibinfo {title} {Structure prediction drives
  materials discovery},}\ }\href@noop {} {\bibfield  {journal} {\bibinfo
  {journal} {Nature Reviews Materials}\ }\textbf {\bibinfo {volume} {4}},\
  \bibinfo {pages} {331--348} (\bibinfo {year} {2019})}\BibitemShut {NoStop}%
\bibitem [{\citenamefont {Segler}\ \emph {et~al.}(2018)\citenamefont {Segler},
  \citenamefont {Preuss},\ and\ \citenamefont {Waller}}]{segler2018planning}%
  \BibitemOpen
  \bibfield  {author} {\bibinfo {author} {\bibfnamefont {Marwin~HS}\
  \bibnamefont {Segler}}, \bibinfo {author} {\bibfnamefont {Mike}\ \bibnamefont
  {Preuss}}, \ and\ \bibinfo {author} {\bibfnamefont {Mark~P}\ \bibnamefont
  {Waller}},\ }\bibfield  {title} {\enquote {\bibinfo {title} {Planning
  chemical syntheses with deep neural networks and symbolic ai},}\ }\href@noop
  {} {\bibfield  {journal} {\bibinfo  {journal} {Nature}\ }\textbf {\bibinfo
  {volume} {555}},\ \bibinfo {pages} {604} (\bibinfo {year}
  {2018})}\BibitemShut {NoStop}%
\bibitem [{\citenamefont {Extance}(2018)}]{extance2018}%
  \BibitemOpen
  \bibfield  {author} {\bibinfo {author} {\bibfnamefont {Andy}\ \bibnamefont
  {Extance}},\ }\bibfield  {title} {\enquote {\bibinfo {title} {How {AI}
  technology can tame the scientific literature},}\ }\href {\doibase
  10.1038/d41586-018-06617-5} {\bibfield  {journal} {\bibinfo  {journal}
  {Nature}\ }\textbf {\bibinfo {volume} {561}},\ \bibinfo {pages} {273--274}
  (\bibinfo {year} {2018})}\BibitemShut {NoStop}%
\bibitem [{\citenamefont {Kahneman}(2016)}]{kahneman2016}%
  \BibitemOpen
  \bibfield  {author} {\bibinfo {author} {\bibfnamefont {Daniel}\ \bibnamefont
  {Kahneman}},\ }\bibfield  {title} {\enquote {\bibinfo {title} {Noise: How to
  overcome the high, hidden cost of inconsistent decision making},}\
  }\href@noop {} {\bibfield  {journal} {\bibinfo  {journal} {Harvard Business
  Review}\ ,\ \bibinfo {pages} {36--43}} (\bibinfo {year} {2016})}\BibitemShut
  {NoStop}%
\bibitem [{\citenamefont {Agrawal}\ \emph {et~al.}(2018)\citenamefont
  {Agrawal}, \citenamefont {Gans},\ and\ \citenamefont
  {Goldfard}}]{agrawal2018prediction}%
  \BibitemOpen
  \bibfield  {author} {\bibinfo {author} {\bibfnamefont {Ajay}\ \bibnamefont
  {Agrawal}}, \bibinfo {author} {\bibfnamefont {Joshua}\ \bibnamefont {Gans}},
  \ and\ \bibinfo {author} {\bibfnamefont {Avi}\ \bibnamefont {Goldfard}},\
  }\href@noop {} {\emph {\bibinfo {title} {Prediction machines: the simple
  economics of artificial intelligence}}}\ (\bibinfo  {publisher} {Harvard
  Business Review Press},\ \bibinfo {address} {Boston},\ \bibinfo {year}
  {2018})\BibitemShut {NoStop}%
\bibitem [{\citenamefont {Rosenfeld}\ \emph {et~al.}(2012)\citenamefont
  {Rosenfeld}, \citenamefont {Zuckerman}, \citenamefont {Azaria},\ and\
  \citenamefont {Kraus}}]{rosenfeld2012combining}%
  \BibitemOpen
  \bibfield  {author} {\bibinfo {author} {\bibfnamefont {Avi}\ \bibnamefont
  {Rosenfeld}}, \bibinfo {author} {\bibfnamefont {Inon}\ \bibnamefont
  {Zuckerman}}, \bibinfo {author} {\bibfnamefont {Amos}\ \bibnamefont
  {Azaria}}, \ and\ \bibinfo {author} {\bibfnamefont {Sarit}\ \bibnamefont
  {Kraus}},\ }\bibfield  {title} {\enquote {\bibinfo {title} {Combining
  psychological models with machine learning to better predict people’s
  decisions},}\ }\href@noop {} {\bibfield  {journal} {\bibinfo  {journal}
  {Synthese}\ }\textbf {\bibinfo {volume} {189}},\ \bibinfo {pages} {81--93}
  (\bibinfo {year} {2012})}\BibitemShut {NoStop}%
\bibitem [{\citenamefont {Risdon}(2017)}]{Risdon2017}%
  \BibitemOpen
  \bibfield  {author} {\bibinfo {author} {\bibfnamefont {Chris}\ \bibnamefont
  {Risdon}},\ }\bibfield  {title} {\enquote {\bibinfo {title} {Scaling nudges
  with machine learning},}\ }\href@noop {} {\bibfield  {journal} {\bibinfo
  {journal} {Behavioral Scientist}\ } (\bibinfo {year} {2017})},\ \bibinfo
  {note} {accessed 2019-04-20}\BibitemShut {NoStop}%
\bibitem [{\citenamefont {Subrahmanian}\ and\ \citenamefont
  {Kumar}(2017)}]{subrahmanian2017predicting}%
  \BibitemOpen
  \bibfield  {author} {\bibinfo {author} {\bibfnamefont {VS}~\bibnamefont
  {Subrahmanian}}\ and\ \bibinfo {author} {\bibfnamefont {Srijan}\ \bibnamefont
  {Kumar}},\ }\bibfield  {title} {\enquote {\bibinfo {title} {Predicting human
  behavior: The next frontiers},}\ }\href@noop {} {\bibfield  {journal}
  {\bibinfo  {journal} {Science}\ }\textbf {\bibinfo {volume} {355}},\ \bibinfo
  {pages} {489--489} (\bibinfo {year} {2017})}\BibitemShut {NoStop}%
\bibitem [{\citenamefont {Ascarza}(2018)}]{ascarza2018retention}%
  \BibitemOpen
  \bibfield  {author} {\bibinfo {author} {\bibfnamefont {Eva}\ \bibnamefont
  {Ascarza}},\ }\bibfield  {title} {\enquote {\bibinfo {title} {Retention
  futility: Targeting high-risk customers might be ineffective},}\ }\href@noop
  {} {\bibfield  {journal} {\bibinfo  {journal} {Journal of Marketing
  Research}\ }\textbf {\bibinfo {volume} {55}},\ \bibinfo {pages} {80--98}
  (\bibinfo {year} {2018})}\BibitemShut {NoStop}%
\bibitem [{\citenamefont {Andini}\ \emph {et~al.}(2018)\citenamefont {Andini},
  \citenamefont {Ciani}, \citenamefont {de~Blasio}, \citenamefont {D'Ignazio},\
  and\ \citenamefont {Salvestrini}}]{andini2018targeting}%
  \BibitemOpen
  \bibfield  {author} {\bibinfo {author} {\bibfnamefont {Monica}\ \bibnamefont
  {Andini}}, \bibinfo {author} {\bibfnamefont {Emanuele}\ \bibnamefont
  {Ciani}}, \bibinfo {author} {\bibfnamefont {Guido}\ \bibnamefont
  {de~Blasio}}, \bibinfo {author} {\bibfnamefont {Alessio}\ \bibnamefont
  {D'Ignazio}}, \ and\ \bibinfo {author} {\bibfnamefont {Viola}\ \bibnamefont
  {Salvestrini}},\ }\bibfield  {title} {\enquote {\bibinfo {title} {Targeting
  with machine learning: An application to a tax rebate program in italy},}\
  }\href@noop {} {\bibfield  {journal} {\bibinfo  {journal} {Journal of
  Economic Behavior \& Organization}\ }\textbf {\bibinfo {volume} {156}},\
  \bibinfo {pages} {86--102} (\bibinfo {year} {2018})}\BibitemShut {NoStop}%
\bibitem [{\citenamefont {Plonsky}\ \emph {et~al.}(2019)\citenamefont
  {Plonsky}, \citenamefont {Apel}, \citenamefont {Ert}, \citenamefont
  {Tennenholtz}, \citenamefont {Bourgin}, \citenamefont {Peterson},
  \citenamefont {Reichman}, \citenamefont {Griffiths}, \citenamefont {Russell},
  \citenamefont {Carter} \emph {et~al.}}]{plonsky2019predicting}%
  \BibitemOpen
  \bibfield  {author} {\bibinfo {author} {\bibfnamefont {Ori}\ \bibnamefont
  {Plonsky}}, \bibinfo {author} {\bibfnamefont {Reut}\ \bibnamefont {Apel}},
  \bibinfo {author} {\bibfnamefont {Eyal}\ \bibnamefont {Ert}}, \bibinfo
  {author} {\bibfnamefont {Moshe}\ \bibnamefont {Tennenholtz}}, \bibinfo
  {author} {\bibfnamefont {David}\ \bibnamefont {Bourgin}}, \bibinfo {author}
  {\bibfnamefont {Joshua~C}\ \bibnamefont {Peterson}}, \bibinfo {author}
  {\bibfnamefont {Daniel}\ \bibnamefont {Reichman}}, \bibinfo {author}
  {\bibfnamefont {Thomas~L}\ \bibnamefont {Griffiths}}, \bibinfo {author}
  {\bibfnamefont {Stuart~J}\ \bibnamefont {Russell}}, \bibinfo {author}
  {\bibfnamefont {Evan~C}\ \bibnamefont {Carter}},  \emph {et~al.},\ }\bibfield
   {title} {\enquote {\bibinfo {title} {Predicting human decisions with
  behavioral theories and machine learning},}\ }\href@noop {} {\bibfield
  {journal} {\bibinfo  {journal} {arXiv e-prints}\ ,\ \bibinfo {eid}
  {arXiv:1904.06866}} (\bibinfo {year} {2019})}\BibitemShut {NoStop}%
\bibitem [{\citenamefont {Loewenstein}\ \emph {et~al.}(2015)\citenamefont
  {Loewenstein}, \citenamefont {Bryce}, \citenamefont {Hagmann},\ and\
  \citenamefont {Rajpal}}]{loewenstein2015}%
  \BibitemOpen
  \bibfield  {author} {\bibinfo {author} {\bibfnamefont {George}\ \bibnamefont
  {Loewenstein}}, \bibinfo {author} {\bibfnamefont {Cindy}\ \bibnamefont
  {Bryce}}, \bibinfo {author} {\bibfnamefont {David}\ \bibnamefont {Hagmann}},
  \ and\ \bibinfo {author} {\bibfnamefont {Sachin}\ \bibnamefont {Rajpal}},\
  }\bibfield  {title} {\enquote {\bibinfo {title} {Warning: You are about to be
  nudged},}\ }\href@noop {} {\bibfield  {journal} {\bibinfo  {journal}
  {Behavioral Science \& Policy}\ }\textbf {\bibinfo {volume} {1}},\ \bibinfo
  {pages} {35 -- 42} (\bibinfo {year} {2015})}\BibitemShut {NoStop}%
\bibitem [{\citenamefont {Mitchell}\ \emph {et~al.}(2018)\citenamefont
  {Mitchell}, \citenamefont {Potash},\ and\ \citenamefont
  {Barocas}}]{mitchell2018prediction}%
  \BibitemOpen
  \bibfield  {author} {\bibinfo {author} {\bibfnamefont {Shira}\ \bibnamefont
  {Mitchell}}, \bibinfo {author} {\bibfnamefont {Eric}\ \bibnamefont {Potash}},
  \ and\ \bibinfo {author} {\bibfnamefont {Solon}\ \bibnamefont {Barocas}},\
  }\bibfield  {title} {\enquote {\bibinfo {title} {Prediction-based decisions
  and fairness: A catalogue of choices, assumptions, and definitions},}\
  }\href@noop {} {\bibfield  {journal} {\bibinfo  {journal} {arXiv preprint}\
  ,\ \bibinfo {eid} {arXiv:1811.07867}} (\bibinfo {year} {2018})}\BibitemShut
  {NoStop}%
\bibitem [{\citenamefont {Athey}(2017)}]{athey2017beyond}%
  \BibitemOpen
  \bibfield  {author} {\bibinfo {author} {\bibfnamefont {Susan}\ \bibnamefont
  {Athey}},\ }\bibfield  {title} {\enquote {\bibinfo {title} {Beyond
  prediction: Using big data for policy problems},}\ }\href@noop {} {\bibfield
  {journal} {\bibinfo  {journal} {Science}\ }\textbf {\bibinfo {volume}
  {355}},\ \bibinfo {pages} {483--485} (\bibinfo {year} {2017})}\BibitemShut
  {NoStop}%
\bibitem [{\citenamefont {Breiman}(2001)}]{breiman2001random}%
  \BibitemOpen
  \bibfield  {author} {\bibinfo {author} {\bibfnamefont {Leo}\ \bibnamefont
  {Breiman}},\ }\bibfield  {title} {\enquote {\bibinfo {title} {Random
  forests},}\ }\href@noop {} {\bibfield  {journal} {\bibinfo  {journal}
  {Machine learning}\ }\textbf {\bibinfo {volume} {45}},\ \bibinfo {pages}
  {5--32} (\bibinfo {year} {2001})}\BibitemShut {NoStop}%
\bibitem [{\citenamefont {Kupferschmidt}(2018)}]{kupferschmidt2018recipe}%
  \BibitemOpen
  \bibfield  {author} {\bibinfo {author} {\bibfnamefont {Kai}\ \bibnamefont
  {Kupferschmidt}},\ }\bibfield  {title} {\enquote {\bibinfo {title} {A recipe
  for rigor},}\ }\href@noop {} {\bibfield  {journal} {\bibinfo  {journal}
  {Science}\ }\textbf {\bibinfo {volume} {361}},\ \bibinfo {pages} {1192--1193}
  (\bibinfo {year} {2018})}\BibitemShut {NoStop}%
\bibitem [{\citenamefont {Hutson}(2018)}]{Hutson2018}%
  \BibitemOpen
  \bibfield  {author} {\bibinfo {author} {\bibfnamefont {Matthew}\ \bibnamefont
  {Hutson}},\ }\bibfield  {title} {\enquote {\bibinfo {title} {Artificial
  intelligence faces reproducibility crisis},}\ }\href@noop {} {\bibfield
  {journal} {\bibinfo  {journal} {Science}\ }\textbf {\bibinfo {volume}
  {359}},\ \bibinfo {pages} {725--726} (\bibinfo {year} {2018})}\BibitemShut
  {NoStop}%
\bibitem [{\citenamefont {Kusner}\ \emph {et~al.}(2017)\citenamefont {Kusner},
  \citenamefont {Loftus}, \citenamefont {Russell},\ and\ \citenamefont
  {Silva}}]{kusner2017counterfactual}%
  \BibitemOpen
  \bibfield  {author} {\bibinfo {author} {\bibfnamefont {Matt~J}\ \bibnamefont
  {Kusner}}, \bibinfo {author} {\bibfnamefont {Joshua}\ \bibnamefont {Loftus}},
  \bibinfo {author} {\bibfnamefont {Chris}\ \bibnamefont {Russell}}, \ and\
  \bibinfo {author} {\bibfnamefont {Ricardo}\ \bibnamefont {Silva}},\
  }\bibfield  {title} {\enquote {\bibinfo {title} {Counterfactual fairness},}\
  }\href@noop {} {\bibfield  {journal} {\bibinfo  {journal} {Advances in Neural
  Information Processing Systems}\ ,\ \bibinfo {pages} {4066--4076}} (\bibinfo
  {year} {2017})}\BibitemShut {NoStop}%
\bibitem [{\citenamefont {Zemel}\ \emph {et~al.}(2013)\citenamefont {Zemel},
  \citenamefont {Wu}, \citenamefont {Swersky}, \citenamefont {Pitassi},\ and\
  \citenamefont {Dwork}}]{zemel2013}%
  \BibitemOpen
  \bibfield  {author} {\bibinfo {author} {\bibfnamefont {Rich}\ \bibnamefont
  {Zemel}}, \bibinfo {author} {\bibfnamefont {Yu}~\bibnamefont {Wu}}, \bibinfo
  {author} {\bibfnamefont {Kevin}\ \bibnamefont {Swersky}}, \bibinfo {author}
  {\bibfnamefont {Toni}\ \bibnamefont {Pitassi}}, \ and\ \bibinfo {author}
  {\bibfnamefont {Cynthia}\ \bibnamefont {Dwork}},\ }\bibfield  {title}
  {\enquote {\bibinfo {title} {Learning fair representations},}\ }\href@noop {}
  {\bibfield  {journal} {\bibinfo  {journal} {Proceedings of International
  Conference on Machine Learning}\ ,\ \bibinfo {pages} {325--333}} (\bibinfo
  {year} {2013})}\BibitemShut {NoStop}%
\bibitem [{\citenamefont {Kleinberg}\ \emph
  {et~al.}(2018{\natexlab{a}})\citenamefont {Kleinberg}, \citenamefont
  {Ludwig}, \citenamefont {Mullainathan},\ and\ \citenamefont
  {Rambachan}}]{Kleinberg2018b}%
  \BibitemOpen
  \bibfield  {author} {\bibinfo {author} {\bibfnamefont {Jon}\ \bibnamefont
  {Kleinberg}}, \bibinfo {author} {\bibfnamefont {Jens}\ \bibnamefont
  {Ludwig}}, \bibinfo {author} {\bibfnamefont {Sendhil}\ \bibnamefont
  {Mullainathan}}, \ and\ \bibinfo {author} {\bibfnamefont {Ashesh}\
  \bibnamefont {Rambachan}},\ }\bibfield  {title} {\enquote {\bibinfo {title}
  {Algorithmic fairness},}\ }\href@noop {} {\bibfield  {journal} {\bibinfo
  {journal} {AEA Papers and Proceedings}\ }\textbf {\bibinfo {volume} {108}},\
  \bibinfo {pages} {22--27} (\bibinfo {year} {2018}{\natexlab{a}})}\BibitemShut
  {NoStop}%
\bibitem [{\citenamefont {Mudrakarta}\ \emph {et~al.}(2018)\citenamefont
  {Mudrakarta}, \citenamefont {Taly}, \citenamefont {Sundararajan},\ and\
  \citenamefont {Dhamdhere}}]{mudrakarta2018did}%
  \BibitemOpen
  \bibfield  {author} {\bibinfo {author} {\bibfnamefont {Pramod~Kaushik}\
  \bibnamefont {Mudrakarta}}, \bibinfo {author} {\bibfnamefont {Ankur}\
  \bibnamefont {Taly}}, \bibinfo {author} {\bibfnamefont {Mukund}\ \bibnamefont
  {Sundararajan}}, \ and\ \bibinfo {author} {\bibfnamefont {Kedar}\
  \bibnamefont {Dhamdhere}},\ }\bibfield  {title} {\enquote {\bibinfo {title}
  {Did the model understand the question?}}\ }\href@noop {} {\bibfield
  {journal} {\bibinfo  {journal} {arXiv e-print}\ ,\ \bibinfo {eid}
  {arXiv:1805.05492}} (\bibinfo {year} {2018})}\BibitemShut {NoStop}%
\bibitem [{\citenamefont {Calmon}\ \emph {et~al.}(2017)\citenamefont {Calmon},
  \citenamefont {Wei}, \citenamefont {Vinzamuri}, \citenamefont
  {Natesan~Ramamurthy},\ and\ \citenamefont {Varshney}}]{calmon2017}%
  \BibitemOpen
  \bibfield  {author} {\bibinfo {author} {\bibfnamefont {Flavio}\ \bibnamefont
  {Calmon}}, \bibinfo {author} {\bibfnamefont {Dennis}\ \bibnamefont {Wei}},
  \bibinfo {author} {\bibfnamefont {Bhanukiran}\ \bibnamefont {Vinzamuri}},
  \bibinfo {author} {\bibfnamefont {Karthikeyan}\ \bibnamefont
  {Natesan~Ramamurthy}}, \ and\ \bibinfo {author} {\bibfnamefont {Kush~R}\
  \bibnamefont {Varshney}},\ }\bibfield  {title} {\enquote {\bibinfo {title}
  {Optimized pre-processing for discrimination prevention},}\ }in\ \href@noop
  {} {\emph {\bibinfo {booktitle} {Advances in Neural Information Processing
  Systems 30}}},\ \bibinfo {editor} {edited by\ \bibinfo {editor}
  {\bibfnamefont {I.}~\bibnamefont {Guyon}}, \bibinfo {editor} {\bibfnamefont
  {U.~V.}\ \bibnamefont {Luxburg}}, \bibinfo {editor} {\bibfnamefont
  {S.}~\bibnamefont {Bengio}}, \bibinfo {editor} {\bibfnamefont
  {H.}~\bibnamefont {Wallach}}, \bibinfo {editor} {\bibfnamefont
  {R.}~\bibnamefont {Fergus}}, \bibinfo {editor} {\bibfnamefont
  {S.}~\bibnamefont {Vishwanathan}}, \ and\ \bibinfo {editor} {\bibfnamefont
  {R.}~\bibnamefont {Garnett}}}\ (\bibinfo  {publisher} {Curran Associates,
  Inc.},\ \bibinfo {year} {2017})\ pp.\ \bibinfo {pages}
  {3992--4001}\BibitemShut {NoStop}%
\bibitem [{\citenamefont {Buchanan}(2019)}]{buchanan2019}%
  \BibitemOpen
  \bibfield  {author} {\bibinfo {author} {\bibfnamefont {Mark}\ \bibnamefont
  {Buchanan}},\ }\bibfield  {title} {\enquote {\bibinfo {title} {The limits of
  machine prediction},}\ }\href@noop {} {\bibfield  {journal} {\bibinfo
  {journal} {Nature Physics}\ }\textbf {\bibinfo {volume} {15}},\ \bibinfo
  {pages} {304} (\bibinfo {year} {2019})}\BibitemShut {NoStop}%
\bibitem [{\citenamefont {Aiello}\ \emph {et~al.}(2019)\citenamefont {Aiello},
  \citenamefont {Schifanella}, \citenamefont {Quercia},\ and\ \citenamefont
  {Del~Prete}}]{aiello2019}%
  \BibitemOpen
  \bibfield  {author} {\bibinfo {author} {\bibfnamefont {Luca~Maria}\
  \bibnamefont {Aiello}}, \bibinfo {author} {\bibfnamefont {Rossano}\
  \bibnamefont {Schifanella}}, \bibinfo {author} {\bibfnamefont {Daniele}\
  \bibnamefont {Quercia}}, \ and\ \bibinfo {author} {\bibfnamefont {Lucia}\
  \bibnamefont {Del~Prete}},\ }\bibfield  {title} {\enquote {\bibinfo {title}
  {Large-scale and high-resolution analysis of food purchases and health
  outcomes},}\ }\href@noop {} {\bibfield  {journal} {\bibinfo  {journal} {EPJ
  Data Science}\ }\textbf {\bibinfo {volume} {8}} (\bibinfo {year}
  {2019})}\BibitemShut {NoStop}%
\bibitem [{\citenamefont {Potash}(2018)}]{potash2018}%
  \BibitemOpen
  \bibfield  {author} {\bibinfo {author} {\bibfnamefont {Eric}\ \bibnamefont
  {Potash}},\ }\bibfield  {title} {\enquote {\bibinfo {title} {Randomization
  bias in field trials to evaluate targeting methods},}\ }\href@noop {}
  {\bibfield  {journal} {\bibinfo  {journal} {Economics Letters}\ }\textbf
  {\bibinfo {volume} {167}},\ \bibinfo {pages} {131--135} (\bibinfo {year}
  {2018})}\BibitemShut {NoStop}%
\bibitem [{\citenamefont {Angwin}\ \emph {et~al.}(2016)\citenamefont {Angwin},
  \citenamefont {Larson}, \citenamefont {Mattu},\ and\ \citenamefont
  {Kirchner}}]{angwin2016}%
  \BibitemOpen
  \bibfield  {author} {\bibinfo {author} {\bibfnamefont {Julia}\ \bibnamefont
  {Angwin}}, \bibinfo {author} {\bibfnamefont {Jeff}\ \bibnamefont {Larson}},
  \bibinfo {author} {\bibfnamefont {Surya}\ \bibnamefont {Mattu}}, \ and\
  \bibinfo {author} {\bibfnamefont {Lauren}\ \bibnamefont {Kirchner}},\
  }\bibfield  {title} {\enquote {\bibinfo {title} {Machine bias},}\ }\href@noop
  {} {\bibfield  {journal} {\bibinfo  {journal} {Pro Publica}\ } (\bibinfo
  {year} {2016})},\ \bibinfo {note} {accessed 2019-06-15}\BibitemShut {NoStop}%
\bibitem [{\citenamefont {Kleinberg}\ \emph
  {et~al.}(2018{\natexlab{b}})\citenamefont {Kleinberg}, \citenamefont
  {Ludwig}, \citenamefont {Mullainathan},\ and\ \citenamefont
  {Sunstein}}]{kleinberg2018}%
  \BibitemOpen
  \bibfield  {author} {\bibinfo {author} {\bibfnamefont {Jon}\ \bibnamefont
  {Kleinberg}}, \bibinfo {author} {\bibfnamefont {Jens}\ \bibnamefont
  {Ludwig}}, \bibinfo {author} {\bibfnamefont {Sendhil}\ \bibnamefont
  {Mullainathan}}, \ and\ \bibinfo {author} {\bibfnamefont {Cass~R}\
  \bibnamefont {Sunstein}},\ }\bibfield  {title} {\enquote {\bibinfo {title}
  {Discrimination in the age of algorithms},}\ }\href@noop {} {\bibfield
  {journal} {\bibinfo  {journal} {Journal of Legal Analysis}\ }\textbf
  {\bibinfo {volume} {10}} (\bibinfo {year} {2018}{\natexlab{b}})}\BibitemShut
  {NoStop}%
\bibitem [{\citenamefont {Silberg}\ and\ \citenamefont
  {Manyika}(2019)}]{silberg2019}%
  \BibitemOpen
  \bibfield  {author} {\bibinfo {author} {\bibfnamefont {Jake}\ \bibnamefont
  {Silberg}}\ and\ \bibinfo {author} {\bibfnamefont {James}\ \bibnamefont
  {Manyika}},\ }\bibfield  {title} {\enquote {\bibinfo {title} {{Notes from the
  AI frontier: Tackling bias in AI (and in humans)}},}\ }\href@noop {}
  {\bibfield  {journal} {\bibinfo  {journal} {McKinsey Global Institute
  (Retrieved from McKinsey online database)}\ } (\bibinfo {year}
  {2019})}\BibitemShut {NoStop}%
\bibitem [{\citenamefont {Kleinberg}\ \emph {et~al.}(2017)\citenamefont
  {Kleinberg}, \citenamefont {Lakkaraju}, \citenamefont {Leskovec},
  \citenamefont {Ludwig},\ and\ \citenamefont {Mullainathan}}]{kleinberg2017}%
  \BibitemOpen
  \bibfield  {author} {\bibinfo {author} {\bibfnamefont {Jon}\ \bibnamefont
  {Kleinberg}}, \bibinfo {author} {\bibfnamefont {Himabindu}\ \bibnamefont
  {Lakkaraju}}, \bibinfo {author} {\bibfnamefont {Jure}\ \bibnamefont
  {Leskovec}}, \bibinfo {author} {\bibfnamefont {Jens}\ \bibnamefont {Ludwig}},
  \ and\ \bibinfo {author} {\bibfnamefont {Sendhil}\ \bibnamefont
  {Mullainathan}},\ }\bibfield  {title} {\enquote {\bibinfo {title} {Human
  decisions and machine predictions},}\ }\href {\doibase 10.1093/qje/qjx032}
  {\bibfield  {journal} {\bibinfo  {journal} {The Quarterly Journal of
  Economics}\ }\textbf {\bibinfo {volume} {133}},\ \bibinfo {pages} {237--293}
  (\bibinfo {year} {2017})}\BibitemShut {NoStop}%
\bibitem [{\citenamefont {Dwork}\ \emph {et~al.}(2018)\citenamefont {Dwork},
  \citenamefont {Immorlica}, \citenamefont {Tauman~Kalai},\ and\ \citenamefont
  {Leiserson}}]{dwork2018}%
  \BibitemOpen
  \bibfield  {author} {\bibinfo {author} {\bibfnamefont {Cynthia}\ \bibnamefont
  {Dwork}}, \bibinfo {author} {\bibfnamefont {Nicole}\ \bibnamefont
  {Immorlica}}, \bibinfo {author} {\bibfnamefont {Adam}\ \bibnamefont
  {Tauman~Kalai}}, \ and\ \bibinfo {author} {\bibfnamefont {Max}\ \bibnamefont
  {Leiserson}},\ }\bibfield  {title} {\enquote {\bibinfo {title} {Decoupled
  classifiers for group-fair and efficient machine learning},}\ }\href@noop {}
  {\bibfield  {journal} {\bibinfo  {journal} {Proceedings of the 1st Conference
  on Fairness, Accountability and Transparency}\ }\bibinfo {series}
  {Proceedings of Machine Learning Research},\ \textbf {\bibinfo {volume}
  {81}},\ \bibinfo {pages} {119--133} (\bibinfo {year} {2018})}\BibitemShut
  {NoStop}%
\bibitem [{\citenamefont {O'Neil}(2016)}]{oneil2016}%
  \BibitemOpen
  \bibfield  {author} {\bibinfo {author} {\bibfnamefont {Cathy}\ \bibnamefont
  {O'Neil}},\ }\href@noop {} {\emph {\bibinfo {title} {Weapons of Math
  Destruction: How Big Data Increases Inequality and Threatens Democracy}}}\
  (\bibinfo  {publisher} {Broadway Books},\ \bibinfo {year} {2016})\BibitemShut
  {NoStop}%
\bibitem [{\citenamefont {Kleinberg}\ \emph {et~al.}(2016)\citenamefont
  {Kleinberg}, \citenamefont {Ludwig},\ and\ \citenamefont
  {Mullainathan}}]{kleinberg2016}%
  \BibitemOpen
  \bibfield  {author} {\bibinfo {author} {\bibfnamefont {Jon}\ \bibnamefont
  {Kleinberg}}, \bibinfo {author} {\bibfnamefont {Jens}\ \bibnamefont
  {Ludwig}}, \ and\ \bibinfo {author} {\bibfnamefont {Sendhil}\ \bibnamefont
  {Mullainathan}},\ }\bibfield  {title} {\enquote {\bibinfo {title} {Solving
  social problems with machine learning},}\ }\href@noop {} {\bibfield
  {journal} {\bibinfo  {journal} {Harvard Business Review}\ } (\bibinfo {year}
  {2016})},\ \bibinfo {note} {accessed 2019-05-19}\BibitemShut {NoStop}%
\bibitem [{\citenamefont {Voosen}(2017)}]{Voosen2017}%
  \BibitemOpen
  \bibfield  {author} {\bibinfo {author} {\bibfnamefont {Paul}\ \bibnamefont
  {Voosen}},\ }\bibfield  {title} {\enquote {\bibinfo {title} {The {AI}
  detectives},}\ }\href@noop {} {\bibfield  {journal} {\bibinfo  {journal}
  {Science}\ }\textbf {\bibinfo {volume} {357}},\ \bibinfo {pages} {22--27}
  (\bibinfo {year} {2017})}\BibitemShut {NoStop}%
\bibitem [{\citenamefont {Ribeiro}\ \emph {et~al.}(2016)\citenamefont
  {Ribeiro}, \citenamefont {Singh},\ and\ \citenamefont
  {Guestrin}}]{ribeiro2016}%
  \BibitemOpen
  \bibfield  {author} {\bibinfo {author} {\bibfnamefont {Marco~Tulio}\
  \bibnamefont {Ribeiro}}, \bibinfo {author} {\bibfnamefont {Sameer}\
  \bibnamefont {Singh}}, \ and\ \bibinfo {author} {\bibfnamefont {Carlos}\
  \bibnamefont {Guestrin}},\ }\href@noop {} {\enquote {\bibinfo {title} {{''Why
  Should I Trust You?'': Explaining the Predictions of Any Classifier}},}\ }
  (\bibinfo {year} {2016})\BibitemShut {NoStop}%
\bibitem [{\citenamefont {Molnar}(2019)}]{molnar2019}%
  \BibitemOpen
  \bibfield  {author} {\bibinfo {author} {\bibfnamefont {Christoph}\
  \bibnamefont {Molnar}},\ }\href@noop {} {\emph {\bibinfo {title}
  {{Interpretable Machine Learning: A Guide for Making Black Box Models
  Explainable}}}}\ (\bibinfo  {publisher} {Christoph Molnar},\ \bibinfo {year}
  {2019})\ \bibinfo {note} {accessed: 2019-05-30}\BibitemShut {NoStop}%
\bibitem [{\citenamefont {Rahwan}\ \emph {et~al.}(2019)\citenamefont {Rahwan},
  \citenamefont {Cebrian}, \citenamefont {Obradovich}, \citenamefont {Bongard},
  \citenamefont {Bonnefon}, \citenamefont {Breazeal}, \citenamefont {Crandall},
  \citenamefont {Christakis}, \citenamefont {Couzin}, \citenamefont {Jackson},
  \citenamefont {Jennings}, \citenamefont {Kamar}, \citenamefont {Kloumann},
  \citenamefont {Larochelle}, \citenamefont {Lazer}, \citenamefont {McElreath},
  \citenamefont {Mislove}, \citenamefont {Parkes}, \citenamefont {Pentland},
  \citenamefont {Roberts}, \citenamefont {Shariff}, \citenamefont {Tenenbaum},\
  and\ \citenamefont {Wellman}}]{Rahwan2019}%
  \BibitemOpen
  \bibfield  {author} {\bibinfo {author} {\bibfnamefont {Iyad}\ \bibnamefont
  {Rahwan}}, \bibinfo {author} {\bibfnamefont {Manuel}\ \bibnamefont
  {Cebrian}}, \bibinfo {author} {\bibfnamefont {Nick}\ \bibnamefont
  {Obradovich}}, \bibinfo {author} {\bibfnamefont {Josh}\ \bibnamefont
  {Bongard}}, \bibinfo {author} {\bibfnamefont {Jean-Fran{\c{c}}ois}\
  \bibnamefont {Bonnefon}}, \bibinfo {author} {\bibfnamefont {Cynthia}\
  \bibnamefont {Breazeal}}, \bibinfo {author} {\bibfnamefont {Jacob~W.}\
  \bibnamefont {Crandall}}, \bibinfo {author} {\bibfnamefont {Nicholas~A.}\
  \bibnamefont {Christakis}}, \bibinfo {author} {\bibfnamefont {Iain~D.}\
  \bibnamefont {Couzin}}, \bibinfo {author} {\bibfnamefont {Matthew~O.}\
  \bibnamefont {Jackson}}, \bibinfo {author} {\bibfnamefont {Nicholas~R.}\
  \bibnamefont {Jennings}}, \bibinfo {author} {\bibfnamefont {Ece}\
  \bibnamefont {Kamar}}, \bibinfo {author} {\bibfnamefont {Isabel~M.}\
  \bibnamefont {Kloumann}}, \bibinfo {author} {\bibfnamefont {Hugo}\
  \bibnamefont {Larochelle}}, \bibinfo {author} {\bibfnamefont {David}\
  \bibnamefont {Lazer}}, \bibinfo {author} {\bibfnamefont {Richard}\
  \bibnamefont {McElreath}}, \bibinfo {author} {\bibfnamefont {Alan}\
  \bibnamefont {Mislove}}, \bibinfo {author} {\bibfnamefont {David~C.}\
  \bibnamefont {Parkes}}, \bibinfo {author} {\bibfnamefont {Alex~'Sandy'}\
  \bibnamefont {Pentland}}, \bibinfo {author} {\bibfnamefont {Margaret~E.}\
  \bibnamefont {Roberts}}, \bibinfo {author} {\bibfnamefont {Azim}\
  \bibnamefont {Shariff}}, \bibinfo {author} {\bibfnamefont {Joshua~B.}\
  \bibnamefont {Tenenbaum}}, \ and\ \bibinfo {author} {\bibfnamefont {Michael}\
  \bibnamefont {Wellman}},\ }\bibfield  {title} {\enquote {\bibinfo {title}
  {Machine behaviour},}\ }\href@noop {} {\bibfield  {journal} {\bibinfo
  {journal} {Nature}\ }\textbf {\bibinfo {volume} {568}},\ \bibinfo {pages}
  {477--486} (\bibinfo {year} {2019})}\BibitemShut {NoStop}%
\bibitem [{\citenamefont {Johansson}\ \emph {et~al.}(2011)\citenamefont
  {Johansson}, \citenamefont {Sonstrod}, \citenamefont {Norinder},\ and\
  \citenamefont {Bostrom}}]{Johansson2011}%
  \BibitemOpen
  \bibfield  {author} {\bibinfo {author} {\bibfnamefont {U.}~\bibnamefont
  {Johansson}}, \bibinfo {author} {\bibfnamefont {C.}~\bibnamefont {Sonstrod}},
  \bibinfo {author} {\bibfnamefont {U.}~\bibnamefont {Norinder}}, \ and\
  \bibinfo {author} {\bibfnamefont {H.}~\bibnamefont {Bostrom}},\ }\bibfield
  {title} {\enquote {\bibinfo {title} {{{T}rade-off between accuracy and
  interpretability for predictive in silico modeling}},}\ }\href@noop {}
  {\bibfield  {journal} {\bibinfo  {journal} {Future Med Chem}\ }\textbf
  {\bibinfo {volume} {3}},\ \bibinfo {pages} {647--663} (\bibinfo {year}
  {2011})}\BibitemShut {NoStop}%
\bibitem [{\citenamefont {Varian}(2016)}]{varian2016causal}%
  \BibitemOpen
  \bibfield  {author} {\bibinfo {author} {\bibfnamefont {Hal~R}\ \bibnamefont
  {Varian}},\ }\bibfield  {title} {\enquote {\bibinfo {title} {Causal inference
  in economics and marketing},}\ }\href@noop {} {\bibfield  {journal} {\bibinfo
   {journal} {Proceedings of the National Academy of Sciences}\ }\textbf
  {\bibinfo {volume} {113}},\ \bibinfo {pages} {7310--7315} (\bibinfo {year}
  {2016})}\BibitemShut {NoStop}%
\bibitem [{\citenamefont {Awad}\ \emph {et~al.}(2018)\citenamefont {Awad},
  \citenamefont {Dsouza}, \citenamefont {Kim}, \citenamefont {Schulz},
  \citenamefont {Henrich}, \citenamefont {Shariff}, \citenamefont {Bonnefon},\
  and\ \citenamefont {Rahwan}}]{awad2018}%
  \BibitemOpen
  \bibfield  {author} {\bibinfo {author} {\bibfnamefont {Edmond}\ \bibnamefont
  {Awad}}, \bibinfo {author} {\bibfnamefont {Sohan}\ \bibnamefont {Dsouza}},
  \bibinfo {author} {\bibfnamefont {Richard}\ \bibnamefont {Kim}}, \bibinfo
  {author} {\bibfnamefont {Jonathan}\ \bibnamefont {Schulz}}, \bibinfo {author}
  {\bibfnamefont {Joseph}\ \bibnamefont {Henrich}}, \bibinfo {author}
  {\bibfnamefont {Azim}\ \bibnamefont {Shariff}}, \bibinfo {author}
  {\bibfnamefont {Jean-Fran{\c{c}}ois}\ \bibnamefont {Bonnefon}}, \ and\
  \bibinfo {author} {\bibfnamefont {Iyad}\ \bibnamefont {Rahwan}},\ }\bibfield
  {title} {\enquote {\bibinfo {title} {The moral machine experiment},}\
  }\href@noop {} {\bibfield  {journal} {\bibinfo  {journal} {Nature}\ }\textbf
  {\bibinfo {volume} {563}},\ \bibinfo {pages} {59} (\bibinfo {year}
  {2018})}\BibitemShut {NoStop}%
\bibitem [{\citenamefont {Frischmann}\ and\ \citenamefont
  {Desai}(2018{\natexlab{a}})}]{frischmann2018}%
  \BibitemOpen
  \bibfield  {author} {\bibinfo {author} {\bibfnamefont {Brett}\ \bibnamefont
  {Frischmann}}\ and\ \bibinfo {author} {\bibfnamefont {Deven}\ \bibnamefont
  {Desai}},\ }\bibfield  {title} {\enquote {\bibinfo {title} {The promise and
  peril of personalization},}\ }\href@noop {} {\bibfield  {journal} {\bibinfo
  {journal} {The Center for Internet and Society}\ } (\bibinfo {year}
  {2018}{\natexlab{a}})}\BibitemShut {NoStop}%
\bibitem [{\citenamefont {Frischmann}\ and\ \citenamefont
  {Desai}(2018{\natexlab{b}})}]{frischmann2018b}%
  \BibitemOpen
  \bibfield  {author} {\bibinfo {author} {\bibfnamefont {Brett}\ \bibnamefont
  {Frischmann}}\ and\ \bibinfo {author} {\bibfnamefont {Deven}\ \bibnamefont
  {Desai}},\ }\bibfield  {title} {\enquote {\bibinfo {title} {How
  personalization leads to homogeneity},}\ }\href@noop {} {\bibfield  {journal}
  {\bibinfo  {journal} {Scientific American}\ } (\bibinfo {year}
  {2018}{\natexlab{b}})}\BibitemShut {NoStop}%
\bibitem [{\citenamefont {Ried}\ \emph {et~al.}(2019)\citenamefont {Ried},
  \citenamefont {M{\"u}ller},\ and\ \citenamefont
  {Briegel}}]{ried2019modelling}%
  \BibitemOpen
  \bibfield  {author} {\bibinfo {author} {\bibfnamefont {Katja}\ \bibnamefont
  {Ried}}, \bibinfo {author} {\bibfnamefont {Thomas}\ \bibnamefont
  {M{\"u}ller}}, \ and\ \bibinfo {author} {\bibfnamefont {Hans~J}\ \bibnamefont
  {Briegel}},\ }\bibfield  {title} {\enquote {\bibinfo {title} {Modelling
  collective motion based on the principle of agency: General framework and the
  case of marching locusts},}\ }\href@noop {} {\bibfield  {journal} {\bibinfo
  {journal} {PloS one}\ }\textbf {\bibinfo {volume} {14}},\ \bibinfo {pages}
  {e0212044} (\bibinfo {year} {2019})}\BibitemShut {NoStop}%
\bibitem [{\citenamefont {Sumpter}(2005)}]{sumpter2005principles}%
  \BibitemOpen
  \bibfield  {author} {\bibinfo {author} {\bibfnamefont {David~JT}\
  \bibnamefont {Sumpter}},\ }\bibfield  {title} {\enquote {\bibinfo {title}
  {The principles of collective animal behaviour},}\ }\href@noop {} {\bibfield
  {journal} {\bibinfo  {journal} {Philosophical Transactions of the Royal
  Society B: Biological Sciences}\ }\textbf {\bibinfo {volume} {361}},\
  \bibinfo {pages} {5--22} (\bibinfo {year} {2005})}\BibitemShut {NoStop}%
\bibitem [{\citenamefont {Schelling}(1978)}]{schelling1978micromotives}%
  \BibitemOpen
  \bibfield  {author} {\bibinfo {author} {\bibfnamefont {Thomas~C}\
  \bibnamefont {Schelling}},\ }\href@noop {} {\emph {\bibinfo {title}
  {Micromotives and Macrobehavior}}}\ (\bibinfo  {publisher} {WW Norton \&
  Company},\ \bibinfo {year} {1978})\BibitemShut {NoStop}%
\bibitem [{\citenamefont {Finlayson}\ \emph {et~al.}(2019)\citenamefont
  {Finlayson}, \citenamefont {Bowers}, \citenamefont {Ito}, \citenamefont
  {Zittrain}, \citenamefont {Beam},\ and\ \citenamefont
  {Kohane}}]{Finlayson1287}%
  \BibitemOpen
  \bibfield  {author} {\bibinfo {author} {\bibfnamefont {Samuel~G.}\
  \bibnamefont {Finlayson}}, \bibinfo {author} {\bibfnamefont {John~D.}\
  \bibnamefont {Bowers}}, \bibinfo {author} {\bibfnamefont {Joichi}\
  \bibnamefont {Ito}}, \bibinfo {author} {\bibfnamefont {Jonathan~L.}\
  \bibnamefont {Zittrain}}, \bibinfo {author} {\bibfnamefont {Andrew~L.}\
  \bibnamefont {Beam}}, \ and\ \bibinfo {author} {\bibfnamefont {Isaac~S.}\
  \bibnamefont {Kohane}},\ }\bibfield  {title} {\enquote {\bibinfo {title}
  {Adversarial attacks on medical machine learning},}\ }\href@noop {}
  {\bibfield  {journal} {\bibinfo  {journal} {Science}\ }\textbf {\bibinfo
  {volume} {363}},\ \bibinfo {pages} {1287--1289} (\bibinfo {year}
  {2019})}\BibitemShut {NoStop}%
\bibitem [{\citenamefont {Kurakin}\ \emph {et~al.}(2016)\citenamefont
  {Kurakin}, \citenamefont {Goodfellow},\ and\ \citenamefont
  {Bengio}}]{Kurakin2016}%
  \BibitemOpen
  \bibfield  {author} {\bibinfo {author} {\bibfnamefont {Alexey}\ \bibnamefont
  {Kurakin}}, \bibinfo {author} {\bibfnamefont {Ian}\ \bibnamefont
  {Goodfellow}}, \ and\ \bibinfo {author} {\bibfnamefont {Samy}\ \bibnamefont
  {Bengio}},\ }\bibfield  {title} {\enquote {\bibinfo {title} {Adversarial
  examples in the physical world},}\ }\href@noop {} {\bibfield  {journal}
  {\bibinfo  {journal} {arXiv e-print}\ ,\ \bibinfo {eid} {arXiv:1607.02533}}
  (\bibinfo {year} {2016})}\BibitemShut {NoStop}%
\bibitem [{\citenamefont {Levitt}\ and\ \citenamefont
  {List}(2007)}]{levitt2007}%
  \BibitemOpen
  \bibfield  {author} {\bibinfo {author} {\bibfnamefont {Steven~D}\
  \bibnamefont {Levitt}}\ and\ \bibinfo {author} {\bibfnamefont {John~A}\
  \bibnamefont {List}},\ }\bibfield  {title} {\enquote {\bibinfo {title} {What
  do laboratory experiments measuring social preferences reveal about the real
  world?}}\ }\href@noop {} {\bibfield  {journal} {\bibinfo  {journal} {Journal
  of Economic Perspectives}\ }\textbf {\bibinfo {volume} {21}},\ \bibinfo
  {pages} {153--174} (\bibinfo {year} {2007})}\BibitemShut {NoStop}%
\bibitem [{\citenamefont {Falk}\ and\ \citenamefont
  {Heckman}(2009)}]{Falk2009}%
  \BibitemOpen
  \bibfield  {author} {\bibinfo {author} {\bibfnamefont {Armin}\ \bibnamefont
  {Falk}}\ and\ \bibinfo {author} {\bibfnamefont {James~J.}\ \bibnamefont
  {Heckman}},\ }\bibfield  {title} {\enquote {\bibinfo {title} {Lab experiments
  are a major source of knowledge in the social sciences},}\ }\href@noop {}
  {\bibfield  {journal} {\bibinfo  {journal} {Science}\ }\textbf {\bibinfo
  {volume} {326}},\ \bibinfo {pages} {535--538} (\bibinfo {year}
  {2009})}\BibitemShut {NoStop}%
\bibitem [{\citenamefont {Charness}\ and\ \citenamefont
  {Fehr}(2015)}]{Charness2015}%
  \BibitemOpen
  \bibfield  {author} {\bibinfo {author} {\bibfnamefont {Gary}\ \bibnamefont
  {Charness}}\ and\ \bibinfo {author} {\bibfnamefont {Ernst}\ \bibnamefont
  {Fehr}},\ }\bibfield  {title} {\enquote {\bibinfo {title} {From the lab to
  the real world},}\ }\href@noop {} {\bibfield  {journal} {\bibinfo  {journal}
  {Science}\ }\textbf {\bibinfo {volume} {350}},\ \bibinfo {pages} {512--513}
  (\bibinfo {year} {2015})}\BibitemShut {NoStop}%
\bibitem [{\citenamefont {Downs}(2013)}]{downs2013}%
  \BibitemOpen
  \bibfield  {author} {\bibinfo {author} {\bibfnamefont {Olly}\ \bibnamefont
  {Downs}},\ }\bibfield  {title} {\enquote {\bibinfo {title} {How data science
  is advancing the “nudge” to influence mobile behaviors},}\ }\href@noop {}
  {\bibfield  {journal} {\bibinfo  {journal} {All Things}\ } (\bibinfo {year}
  {2013})},\ \bibinfo {note} {accessed 2019-05-19}\BibitemShut {NoStop}%
\bibitem [{\citenamefont {Milkman}\ \emph {et~al.}(2011)\citenamefont
  {Milkman}, \citenamefont {Beshears}, \citenamefont {Choi}, \citenamefont
  {Laibson},\ and\ \citenamefont {Madrian}}]{Milkman2011}%
  \BibitemOpen
  \bibfield  {author} {\bibinfo {author} {\bibfnamefont {Katherine~L}\
  \bibnamefont {Milkman}}, \bibinfo {author} {\bibfnamefont {John}\
  \bibnamefont {Beshears}}, \bibinfo {author} {\bibfnamefont {James~J}\
  \bibnamefont {Choi}}, \bibinfo {author} {\bibfnamefont {David}\ \bibnamefont
  {Laibson}}, \ and\ \bibinfo {author} {\bibfnamefont {Brigitte~C}\
  \bibnamefont {Madrian}},\ }\bibfield  {title} {\enquote {\bibinfo {title}
  {Using implementation intentions prompts to enhance influenza vaccination
  rates},}\ }\href@noop {} {\bibfield  {journal} {\bibinfo  {journal}
  {Proceedings of the National Academy of Sciences}\ }\textbf {\bibinfo
  {volume} {108}},\ \bibinfo {pages} {10415--10420} (\bibinfo {year}
  {2011})}\BibitemShut {NoStop}%
\bibitem [{\citenamefont {Zeevi}\ \emph {et~al.}(2015)\citenamefont {Zeevi},
  \citenamefont {Korem}, \citenamefont {Zmora}, \citenamefont {Israeli},
  \citenamefont {Rothschild}, \citenamefont {Weinberger}, \citenamefont
  {Ben-Yacov}, \citenamefont {Lador}, \citenamefont {Avnit-Sagi}, \citenamefont
  {Lotan-Pompan}, \citenamefont {Suez}, \citenamefont {Mahdi}, \citenamefont
  {Matot}, \citenamefont {Malka}, \citenamefont {Kosower}, \citenamefont
  {Rein}, \citenamefont {Zilberman-Schapira}, \citenamefont {Dohnalová},
  \citenamefont {Pevsner-Fischer}, \citenamefont {Bikovsky}, \citenamefont
  {Halpern}, \citenamefont {Elinav},\ and\ \citenamefont {Segal}}]{Zeevi2015}%
  \BibitemOpen
  \bibfield  {author} {\bibinfo {author} {\bibfnamefont {David}\ \bibnamefont
  {Zeevi}}, \bibinfo {author} {\bibfnamefont {Tal}\ \bibnamefont {Korem}},
  \bibinfo {author} {\bibfnamefont {Niv}\ \bibnamefont {Zmora}}, \bibinfo
  {author} {\bibfnamefont {David}\ \bibnamefont {Israeli}}, \bibinfo {author}
  {\bibfnamefont {Daphna}\ \bibnamefont {Rothschild}}, \bibinfo {author}
  {\bibfnamefont {Adina}\ \bibnamefont {Weinberger}}, \bibinfo {author}
  {\bibfnamefont {Orly}\ \bibnamefont {Ben-Yacov}}, \bibinfo {author}
  {\bibfnamefont {Dar}\ \bibnamefont {Lador}}, \bibinfo {author} {\bibfnamefont
  {Tali}\ \bibnamefont {Avnit-Sagi}}, \bibinfo {author} {\bibfnamefont {Maya}\
  \bibnamefont {Lotan-Pompan}}, \bibinfo {author} {\bibfnamefont {Jotham}\
  \bibnamefont {Suez}}, \bibinfo {author} {\bibfnamefont {Jemal~Ali}\
  \bibnamefont {Mahdi}}, \bibinfo {author} {\bibfnamefont {Elad}\ \bibnamefont
  {Matot}}, \bibinfo {author} {\bibfnamefont {Gal}\ \bibnamefont {Malka}},
  \bibinfo {author} {\bibfnamefont {Noa}\ \bibnamefont {Kosower}}, \bibinfo
  {author} {\bibfnamefont {Michal}\ \bibnamefont {Rein}}, \bibinfo {author}
  {\bibfnamefont {Gili}\ \bibnamefont {Zilberman-Schapira}}, \bibinfo {author}
  {\bibfnamefont {Lenka}\ \bibnamefont {Dohnalová}}, \bibinfo {author}
  {\bibfnamefont {Meirav}\ \bibnamefont {Pevsner-Fischer}}, \bibinfo {author}
  {\bibfnamefont {Rony}\ \bibnamefont {Bikovsky}}, \bibinfo {author}
  {\bibfnamefont {Zamir}\ \bibnamefont {Halpern}}, \bibinfo {author}
  {\bibfnamefont {Eran}\ \bibnamefont {Elinav}}, \ and\ \bibinfo {author}
  {\bibfnamefont {Eran}\ \bibnamefont {Segal}},\ }\bibfield  {title} {\enquote
  {\bibinfo {title} {Personalized nutrition by prediction of glycemic
  responses},}\ }\href@noop {} {\bibfield  {journal} {\bibinfo  {journal}
  {Cell}\ }\textbf {\bibinfo {volume} {163}},\ \bibinfo {pages} {1079 -- 1094}
  (\bibinfo {year} {2015})}\BibitemShut {NoStop}%
\bibitem [{\citenamefont {Park}\ \emph {et~al.}(2019)\citenamefont {Park},
  \citenamefont {Reiner}, \citenamefont {Green},\ and\ \citenamefont
  {Williamson}}]{Park2019}%
  \BibitemOpen
  \bibfield  {author} {\bibinfo {author} {\bibfnamefont {Toby}\ \bibnamefont
  {Park}}, \bibinfo {author} {\bibfnamefont {Carolin}\ \bibnamefont {Reiner}},
  \bibinfo {author} {\bibfnamefont {Kevin}\ \bibnamefont {Green}}, \ and\
  \bibinfo {author} {\bibfnamefont {Katie}\ \bibnamefont {Williamson}},\
  }\bibfield  {title} {\enquote {\bibinfo {title} {Behavior change for nature:
  A behavioral science toolkit for practitioners},}\ }\href@noop {} {\bibfield
  {journal} {\bibinfo  {journal} {Rare and The Behavioural Insights Team}\ }
  (\bibinfo {year} {2019})}\BibitemShut {NoStop}%
\bibitem [{\citenamefont {Thaler}\ and\ \citenamefont
  {Sunstein}(2003)}]{thaler2003}%
  \BibitemOpen
  \bibfield  {author} {\bibinfo {author} {\bibfnamefont {Richard~H}\
  \bibnamefont {Thaler}}\ and\ \bibinfo {author} {\bibfnamefont {Cass~R}\
  \bibnamefont {Sunstein}},\ }\bibfield  {title} {\enquote {\bibinfo {title}
  {Libertarian paternalism},}\ }\href@noop {} {\bibfield  {journal} {\bibinfo
  {journal} {American Economic Review}\ }\textbf {\bibinfo {volume} {93}},\
  \bibinfo {pages} {175--179} (\bibinfo {year} {2003})}\BibitemShut {NoStop}%
\bibitem [{\citenamefont {Camerer}\ \emph {et~al.}(2003)\citenamefont
  {Camerer}, \citenamefont {Issacharoff}, \citenamefont {Loewenstein},
  \citenamefont {O'Donoghue},\ and\ \citenamefont {Rabin}}]{camerer2003}%
  \BibitemOpen
  \bibfield  {author} {\bibinfo {author} {\bibfnamefont {Colin}\ \bibnamefont
  {Camerer}}, \bibinfo {author} {\bibfnamefont {Samuel}\ \bibnamefont
  {Issacharoff}}, \bibinfo {author} {\bibfnamefont {George}\ \bibnamefont
  {Loewenstein}}, \bibinfo {author} {\bibfnamefont {Ted}\ \bibnamefont
  {O'Donoghue}}, \ and\ \bibinfo {author} {\bibfnamefont {Matthew}\
  \bibnamefont {Rabin}},\ }\bibfield  {title} {\enquote {\bibinfo {title}
  {{Regulation for Conservatives: Behavioral Economics and the Case for
  'Asymmetric Paternalism'}},}\ }\href@noop {} {\bibfield  {journal} {\bibinfo
  {journal} {University of Pennsylvania law review}\ }\textbf {\bibinfo
  {volume} {151}},\ \bibinfo {pages} {1211--1254} (\bibinfo {year}
  {2003})}\BibitemShut {NoStop}%
\bibitem [{\citenamefont {Bresnahan}\ and\ \citenamefont
  {Trajtenberg}(1992)}]{bresnahan1992}%
  \BibitemOpen
  \bibfield  {author} {\bibinfo {author} {\bibfnamefont {Timothy~F}\
  \bibnamefont {Bresnahan}}\ and\ \bibinfo {author} {\bibfnamefont {Manuel}\
  \bibnamefont {Trajtenberg}},\ }\href@noop {} {\emph {\bibinfo {title}
  {General Purpose Technologies. Engines of Growth?}}},\ \bibinfo {type}
  {Working Paper}\ \bibinfo {number} {4148}\ (\bibinfo  {institution} {National
  Bureau of Economic Research},\ \bibinfo {year} {1992})\BibitemShut {NoStop}%
\end{thebibliography}
%

\end{document}